\documentclass[apjl]{emulateapj}
\usepackage{comment}
\usepackage{ifthen}

\newcommand{\forloop}[5][1]%
{%
\setcounter{#2}{#3}%
\ifthenelse{#4}%
	{%
	#5%
	\addtocounter{#2}{#1}%
	\forloop[#1]{#2}{\value{#2}}{#4}{#5}%
	}%
	{%
	}%
}%

\newcommand{\ctbd}[1]{}

\newcommand{\lc}{light curve}

\newcommand{\Lc}{Light curve}

\newcommand{\band}[1]{\ensuremath{#1}~band}

\newcommand{\kms}{\ensuremath{\rm km\,s^{-1}}}
\newcommand{\ms}{\ensuremath{\rm m\,s^{-1}}}

\newcommand{\gcmc}{\ensuremath{\rm g\,cm^{-3}}}
\newcommand{\ergscmsq}{\ensuremath{\rm erg\,s^{-1}\,cm^{-2}}}

\newcommand{\msini}{\ensuremath{m \sin i}}

\newcommand{\vsini}{\ensuremath{v \sin{i}}}
\newcommand{\feh}{\ensuremath{\rm [Fe/H]}}

\newcommand{\rsun}{\ensuremath{R_\sun}}
\newcommand{\msun}{\ensuremath{M_\sun}}
\newcommand{\lsun}{\ensuremath{L_\sun}}

\newcommand{\rstar}{\ensuremath{R_\star}}
\newcommand{\mstar}{\ensuremath{M_\star}}
\newcommand{\lstar}{\ensuremath{L_\star}}

\newcommand{\teffstar}{\ensuremath{T_{\rm eff\star}}}
\newcommand{\rhostar}{\ensuremath{\rho_\star}}
\newcommand{\loggstar}{\ensuremath{\log{g_{\star}}}}

\newcommand{\mearth}{\ensuremath{M_\earth}}

\newcommand{\rpl}{\ensuremath{R_{p}}}
\newcommand{\mpl}{\ensuremath{M_{p}}}

\newcommand{\rhopl}{\ensuremath{\rho_{p}}}

\newcommand{\arstar}{\ensuremath{a/\rstar}}
\newcommand{\zrstar}{\ensuremath{\zeta/\rstar}}

\newcommand{\rjup}{\ensuremath{R_{\rm J}}}
\newcommand{\mjup}{\ensuremath{M_{\rm J}}}

\newcommand{\refsecl}[1]{\mbox{Section \ref{sec:#1}}}

\newcommand{\reftabl}[1]{Table~\ref{tab:#1}}

\newcommand{\hatcurCCra}{\ensuremath{11^{\mathrm h}21^{\mathrm m}18.00{\mathrm s}}}                     
\newcommand{\hatcurCCdec}{\ensuremath{-22{\arcdeg}23{\arcmin}17.4{\arcsec}}}                    
\newcommand{\hatcurCCtwomass}{2MASS~11211786-2223174}     
\newcommand{\hatcurCCgsc}{GSC~6090-00133}                 
\newcommand{\hatcurCCtassmv}{\ensuremath{13.951\pm0.030}} 
\newcommand{\hatcurCCtassmB}{\ensuremath{14.727\pm0.020}} 
\newcommand{\hatcurCCtassmg}{\ensuremath{14.286\pm0.030}} 
\newcommand{\hatcurCCtassmr}{\ensuremath{13.725\pm0.030}} 
\newcommand{\hatcurCCtassmi}{\ensuremath{13.551\pm0.040}} 
\newcommand{\hatcurCCtwomassJmag}{\ensuremath{12.590\pm0.024}} 
\newcommand{\hatcurCCtwomassHmag}{\ensuremath{12.299\pm0.030}} 
\newcommand{\hatcurCCtwomassKmag}{\ensuremath{12.238\pm0.030}} 
\newcommand{\hatcurLCrprstar}{\ensuremath{0.1116\pm0.0021}} 
\newcommand{\hatcurLCbsq}{\ensuremath{0.209_{-0.056}^{+0.054}}} 
\newcommand{\hatcurLCimp}{\ensuremath{0.457_{-0.066}^{+0.056}}} 
\newcommand{\hatcurLCzeta}{\ensuremath{15.23\pm0.13}}     
\newcommand{\hatcurLCdur}{\ensuremath{0.1497\pm0.0017}}   
\newcommand{\hatcurLCingdur}{\ensuremath{0.0186\pm0.0016}} 
\newcommand{\hatcurLCP}{\ensuremath{5.416081\pm0.000016}} 
\newcommand{\hatcurLCPprec}{\ensuremath{5.4160810}}       
\newcommand{\hatcurLCPshort}{\ensuremath{5.4161}}         
\newcommand{\hatcurLCT}{\ensuremath{2456620.66527\pm0.00052}} 
\newcommand{\hatcurLCrho}{\ensuremath{1.59\pm0.54}}       
\newcommand{\hatcurSMEiteff}{\ensuremath{5838\pm89}}      
\newcommand{\hatcurSMEizfeh}{\ensuremath{0.260\pm0.052}}  
\newcommand{\hatcurSMEizfehshort}{\ensuremath{0.26}}      
\newcommand{\hatcurSMEilogg}{\ensuremath{4.60\pm0.12}}    
\newcommand{\hatcurSMEivsin}{\ensuremath{2.42\pm0.75}}    
\newcommand{\hatcurSMEivmac}{\ensuremath{0.0}}            
\newcommand{\hatcurSMEivmic}{\ensuremath{0.0}}            
\newcommand{\hatcurSMEiiteff}{\ensuremath{5670\pm91}}     
\newcommand{\hatcurSMEiizfeh}{\ensuremath{0.180\pm0.064}} 
\newcommand{\hatcurSMEiizfehshort}{\ensuremath{0.18}}     
\newcommand{\hatcurSMEiilogg}{\ensuremath{4.415\pm0.046}} 
\newcommand{\hatcurSMEiivsin}{\ensuremath{2.80\pm0.61}}   
\newcommand{\hatcurLBiz}{\ensuremath{0.2259}}             
\newcommand{\hatcurLBiiz}{\ensuremath{0.3232}}            
\newcommand{\hatcurLBii}{\ensuremath{0.2930}}             
\newcommand{\hatcurLBiii}{\ensuremath{0.3208}}            
\newcommand{\hatcurLBig}{\ensuremath{0.5965}}             
\newcommand{\hatcurLBiig}{\ensuremath{0.2045}}            
\newcommand{\hatcurLBir}{\ensuremath{0.3896}}             
\newcommand{\hatcurLBiir}{\ensuremath{0.3085}}            
\newcommand{\hatcurLBiR}{\ensuremath{0.3628}}             
\newcommand{\hatcurLBiiR}{\ensuremath{0.3129}}            
\newcommand{\hatcurISOm}{\ensuremath{1.038\pm0.039}}      
\newcommand{\hatcurISOmlong}{\ensuremath{1.038\pm0.039}}  
\newcommand{\hatcurISOr}{\ensuremath{1.036\pm0.067}}      
\newcommand{\hatcurISOrlong}{\ensuremath{1.036\pm0.067}}  
\newcommand{\hatcurISOrho}{\ensuremath{1.31\pm0.24}}      
\newcommand{\hatcurISOlogg}{\ensuremath{4.422\pm0.053}}   
\newcommand{\hatcurISOlum}{\ensuremath{0.99\pm0.16}}      
\newcommand{\hatcurISOmv}{\ensuremath{4.86\pm0.19}}       
\newcommand{\hatcurISOage}{\ensuremath{4.3\pm2.3}}        
\newcommand{\hatcurISOMK}{\ensuremath{3.24\pm0.14}}       
\newcommand{\hatcurRVK}{\ensuremath{92.1\pm7.8}}          
\newcommand{\hatcurRVrk}{\ensuremath{-0.233\pm0.084}}     
\newcommand{\hatcurRVrh}{\ensuremath{-0.25_{-0.11}^{+0.18}}} 
\newcommand{\hatcurRVk}{\ensuremath{-0.082\pm0.034}}      
\newcommand{\hatcurRVh}{\ensuremath{-0.090\pm0.065}}      
\newcommand{\hatcurRVgammaA}{\ensuremath{-10887\pm11}}    
\newcommand{\hatcurRVjittertwosiglimA}{\ensuremath{<20.7}} 
\newcommand{\hatcurRVjitterB}{\ensuremath{58\pm44}}       
\newcommand{\hatcurRVjitterC}{\ensuremath{24\pm14}}       
\newcommand{\hatcurRVjitterD}{\ensuremath{93\pm40}}       
\newcommand{\hatcurRVeccen}{\ensuremath{0.129\pm0.049}}   
\newcommand{\hatcurRVomega}{\ensuremath{227\pm29}}        
\newcommand{\hatcurPPi}{\ensuremath{88.10\pm0.33}}        
\newcommand{\hatcurPPlogg}{\ensuremath{3.195\pm0.069}}    
\newcommand{\hatcurPPar}{\ensuremath{12.66\pm0.77}}       
\newcommand{\hatcurPParel}{\ensuremath{0.06112\pm0.00076}} 
\newcommand{\hatcurPPrho}{\ensuremath{0.70\pm0.16}}       
\newcommand{\hatcurPPmlong}{\ensuremath{0.806\pm0.069}}   
\newcommand{\hatcurPPrlong}{\ensuremath{1.126\pm0.077}}   
\newcommand{\hatcurPPmrcorr}{\ensuremath{0.05}}           
\newcommand{\hatcurPPteff}{\ensuremath{1128\pm40}}        
\newcommand{\hatcurPPtheta}{\ensuremath{0.0841\pm0.0093}} 
\newcommand{\hatcurPPfluxavg}{\ensuremath{3.66\pm0.53}}   
\newcommand{\hatcurPPfluxavgdim}{\ensuremath{8}}          
\newcommand{\hatcurcLCdur}{\ensuremath{0.957\pm0.054}}    
\newcommand{\hatcurcLCingdur}{\ensuremath{0.0863\pm0.0011}} 
\newcommand{\hatcurcLCP}{\ensuremath{1422\pm14}}          
\newcommand{\hatcurcLCPshort}{\ensuremath{1422}}          
\newcommand{\hatcurcLCT}{\ensuremath{2456521\pm11}}       
\newcommand{\hatcurcRVK}{\ensuremath{224\pm14}}           
\newcommand{\hatcurcPPar}{\ensuremath{518\pm32}}          
\newcommand{\hatcurcPParel}{\ensuremath{2.504\pm0.035}}   
\newcommand{\hatcurcPPm}{\ensuremath{12.70\pm0.87}}       
\newcommand{\hatcurcPPmlong}{\ensuremath{12.70\pm0.87}}   
\newcommand{\hatcurcPPteff}{\ensuremath{175.9\pm6.4}}     
\newcommand{\hatcurcPPfluxavg}{\ensuremath{2.16\pm0.32}}  
\newcommand{\hatcurcPPfluxavgdim}{\ensuremath{5}}         
\newcommand{\hatcurXAv}{\ensuremath{0.091\pm0.074}}       
\newcommand{\hatcurXdistred}{\ensuremath{630\pm43}}       
\newcommand{\hatcurcRVeccentwosiglimecceneccen}{\ensuremath{<0.083}} 
\newcommand{\hatcur}{HATS-59}
\newcommand{\hatcurb}{HATS-59b}
\newcommand{\hatcurc}{HATS-59c}
\newcommand{\hatcurbc}{HATS-59b,c}
\newcommand{\hatcurRVgammaabs}{\hatcurRVgammaA}                           

\newcommand{\hatcurisocite}{yi:2001}
\newcommand{\hatcurlumind}{\rhostar}
\newcommand{\hatcurjhkfilset}{ESO}

\newcommand{\hatcurSMEversion}{ii}                                       
\newcommand{\hatcurSMEteff}{\ifthenelse{\equal{\hatcurSMEversion}{i}}{\hatcurSMEiteff}{\hatcurSMEiiteff}}
\newcommand{\hatcurSMEzfeh}{\ifthenelse{\equal{\hatcurSMEversion}{i}}{\hatcurSMEizfeh}{\hatcurSMEiizfeh}}
\newcommand{\hatcurSMEzfehshort}{\ifthenelse{\equal{\hatcurSMEversion}{i}}{\hatcurSMEizfehshort}{\hatcurSMEiizfehshort}}
\newcommand{\hatcurSMElogg}{\ifthenelse{\equal{\hatcurSMEversion}{i}}{\hatcurSMEilogg}{\hatcurSMEiilogg}}
\newcommand{\hatcurSMEvsin}{\ifthenelse{\equal{\hatcurSMEversion}{i}}{\hatcurSMEivsin}{\hatcurSMEiivsin}}
\newcommand{\hatcurSMEvmac}{\ifthenelse{\equal{\hatcurSMEversion}{i}}{\hatcurSMEivmac}{\hatcurSMEiivmac}}
\newcommand{\hatcurSMEvmic}{\ifthenelse{\equal{\hatcurSMEversion}{i}}{\hatcurSMEivmic}{\hatcurSMEiivmic}}

\newboolean{emulateapj}
\setboolean{emulateapj}{true}

\newboolean{rvtablelong}
\setboolean{rvtablelong}{true}

\newboolean{astroph}
\setboolean{astroph}{true}

\shortauthors{Sarkis et al.}
\shorttitle{
\hatcur\lowercase{b}
}
\ifthenelse{\boolean{emulateapj}}{
    \newcommand{\titledag}{$\dagger$}
}{
    \newcommand{\titledag}{\dagger}
}

\begin{document}
\title{
\hatcur\lowercase{b,c}: A Transiting Hot Jupiter and a Cold Massive Giant Planet Around a Sun-Like Star
\altaffilmark{\titledag}
}

\label{authorList}
\author{
    P.~Sarkis\altaffilmark{1},
	Th.~Henning\altaffilmark{1},
	J.~D.~Hartman\altaffilmark{2}, 
    G.~\'A.~Bakos\altaffilmark{2,3$\star$}, 
	R.~Brahm\altaffilmark{4,5}, 
	A.~Jord\'an\altaffilmark{1,4,5}, 
	D.~Bayliss\altaffilmark{6}, 	
	L.~Mancini\altaffilmark{7,1,8},	
	N.~Espinoza\altaffilmark{1,4,5}, 
	M.~Rabus\altaffilmark{1,4}, 
	Z.~Csubry\altaffilmark{2},  
	W.~Bhatti\altaffilmark{2}, 
	K.~Penev\altaffilmark{9}, 
	G.~Zhou\altaffilmark{10},  
	J.~Bento\altaffilmark{11}, 
	T.~G.~Tan\altaffilmark{12}, 
    P.~Arriagada\altaffilmark{13}, 
	R.~P.~Butler\altaffilmark{14}, 
	J.~D.~Crane\altaffilmark{14}, 
	S.~Shectman\altaffilmark{14}, 
	C.~G.~Tinney\altaffilmark{15,16}, 
	D.~J.~Wright\altaffilmark{15,16}, 
	B.~Addison\altaffilmark{17}, 
	S.~Durkan\altaffilmark{18}, 
	V.~Suc\altaffilmark{4}, 
	L.~A.~Buchhave\altaffilmark{19}, 
	M.~de~Val-Borro\altaffilmark{20}, 
	J.~L\'az\'ar\altaffilmark{21}, 
	I.~Papp\altaffilmark{21}, 
	P.~S\'ari\altaffilmark{21} 
}
\altaffiltext{1}{Max Planck Institute for Astronomy, Heidelberg, Germany; sarkis@mpia.de}
\altaffiltext{2}{Department of Astrophysical Sciences, Princeton University, NJ 08544, USA} 
\altaffiltext{3}{MTA Distinguished Guest Fellow, Konkoly Observatory}
\altaffiltext{4}{Instituto de Astrof\'isica, Facultad de F\'isica,
  Pontificia Universidad Cat\'olica de Chile, Av. Vicu\~na Mackenna
  4860, 7820436 Macul, Santiago, Chile} 
\altaffiltext{5}{Millennium Institute of Astrophysics, Av. Vicu\~na Mackenna 4860, 7820436 Macul, Santiago, Chile} 
\altaffiltext{6}{Department of Physics, University of Warwick, Coventry CV4 7AL, UK}
\altaffiltext{7}{Department of Physics, University of Rome Tor Vergata, Via della Ricerca Scientifica 1, I-00133 Rome, Italy} 
\altaffiltext{8}{INAF--Astrophysical Observatory of Turin, via Osservatorio 20, I-10025 Pino Torinese, Italy} 
\altaffiltext{9}{Physics Department, University of Texas at Dallas, 800 W Campbell Rd. MS WT15, Richardson, TX 75080, USA} 
\altaffiltext{10}{Harvard-Smithsonian Center for Astrophysics,
  Cambridge, MA 02138, USA} 
\altaffiltext{11}{Research School of Astronomy and Astrophysics, Australian National University, Canberra, ACT 2611, Australia} 
\altaffiltext{12}{Perth Exoplanet Survey Telescope, Perth, Australia} 
\altaffiltext{13}{Department of Terrestrial Magnetism, Carnegie Institution for Science, Washington, DC 20015, USA} 
\altaffiltext{14}{The Observatories of the Carnegie Institution for Science, 813 Santa Barbara St, Pasadena, CA 91101, USA} 
\altaffiltext{15}{Australian Centre for Astrobiology, School of Physics, University of New South Wales, NSW 2052, Australia} 
\altaffiltext{16}{Exoplanetary Science at UNSW, School of Physics, University of New South Wales, NSW 2052, Australia} 
\altaffiltext{17}{Mississippi State University, Department of Physics \& Astronomy, Hilbun Hall, Starkville, MS 39762, USA}  
\altaffiltext{18}{Astrophysics Research Centre, Queens University, Belfast, Belfast, Northern Ireland, UK} 
\altaffiltext{19}{DTU Space, National Space Institute, Technical University of Denmark, Elektrovej 328, DK-2800 Kgs. Lyngby, Denmark} 
\altaffiltext{20}{Astrochemistry Laboratory, Goddard Space Flight Center, NASA, 8800 Greenbelt Rd, Greenbelt, MD 20771, USA} 
\altaffiltext{21}{Hungarian Astronomical Association, 1451 Budapest, Hungary} 
\altaffiltext{$\star$}{Packard Fellow}
\altaffiltext{$\dagger$}{
The HATSouth network is operated by a collaboration consisting of
Princeton University (PU), the Max Planck Institute f\"ur Astronomie
(MPIA), the Australian National University (ANU), and the Pontificia
Universidad Cat\'olica de Chile (PUC).  The station at Las Campanas
Observatory (LCO) of the Carnegie Institute is operated by PU in
conjunction with PUC, the station at the High Energy Spectroscopic
Survey (H.E.S.S.) site is operated in conjunction with MPIA, and the
station at Siding Spring Observatory (SSO) is operated jointly with
ANU.
This paper includes data gathered with
6.5\,m Magellan Telescopes
located as Las Campanas Observatory, Chile,
and
the MPG~2.2\,m, the NTT, and the Euler 1.2\,m
telescopes at the ESO Observatory in La Silla.
This paper uses observations obtained with facilities of the Las Cumbres
Observatory Global Telescope.
Based in part on observations made with
the 3.9\,m Anglo-Australian Telescope and the ANU 2.3\,m Telescope
both at SSO.
Based in part on observations made with
the facilities of the Las Cumbres Observatory Global Telescope,
the Perth Exoplanet Survey Telescope,
and the Nordic Optical Telescope.
}


\label{abstract}
\begin{abstract}

\setcounter{footnote}{10}
We report the first discovery 
of a multi-planetary system 
by the HATSouth network, \hatcurbc{}, 
a planetary system 
with an inner transiting hot Jupiter
and an outer cold massive giant planet,
which was detected via radial velocity.
The inner transiting planet, \hatcurb{},
is on an eccentric orbit with $e = \hatcurRVeccen$,
orbiting a $V=\hatcurCCtassmv$\, mag solar-like star
($\mstar$ = \hatcurISOm\,\msun\, and $\rstar = \hatcurISOr\,\rsun$)
with a period of \hatcurLCP\,days.
The outer companion, \hatcurc{} 
is on a circular orbit
with
$\msini = \hatcurcPPm{}$\,\mjup\, and 
a period of $\hatcurcLCP$\,days.
The inner planet has a mass of
\hatcurPPmlong\,\mjup\, and a 
radius of \hatcurPPrlong\,\rjup\,,
yielding a density of $\hatcurPPrho{}$ \gcmc.
Unlike most of the planetary systems
that include only a single hot Jupiter,
\hatcurbc{} includes, 
in addition to the transiting hot Jupiter,
a massive outer companion.
The architecture of this system is valuable
for understanding planet migration.
\setcounter{footnote}{0}
\end{abstract}



\section{Introduction}
\label{sec:introduction}
During the past decade,
the number of exoplanets has increased steadily 
and by now more than 3500 exoplanets have been statistically validated.
Exoplanets are very common and have a wide variety of properties
\citep[for a review check][]{winn:2015},
which offer a unique opportunity to constrain 
their formation and evolution \citep{Mordasini:2016,Jin:2018}.
Hot Jupiters, i.e. gas giant planets on short orbital periods,
still pose many challenges for planet formation models.
It is believed that such planets formed beyond the iceline,
several au from the central star, 
and migrated inwards through interactions with the disk \citep[e.g.][]{Lin:1996}.
However, disk migration predicts circular and aligned orbits \citep[e.g.][]{GT:1980,art:1993}
and cannot explain the existence of several
hot Jupiters that have been found on retrogade or misaligned orbits 
\citep[for a review see ][]{winn:2015}.
Alternative scenarios have been thus proposed,
which involve interactions 
with a third distant body or planet-planet scattering that can result
in eccentric and misaligned orbits
\citep{kozai:62,lidov:62,nagasawa:2008,li:2014,petrovich:2015}.

One approach to put constraints 
on the different migration mechanisms
is to measure the spin-orbit alignment 
via the Rossiter-McLaughlin effect
\citep[e.g.,][]{Queloz:2000,Zhou:2015}.
Another approach
is to search for planetary or stellar companions
at large separations,
which could have influenced the dynamical evolution
of the inner planet.
\cite{Knutson:2014} performed a long term radial velocity 
monitoring of 51 systems known to host a hot Jupiter,
with the goal to detect further planetary companions.
They estimated an occurrence rate of $51 \pm 10$\%
for companions with masses between 1-13$M_\mathrm{J}$
and orbital semi-major axes between 1-20 au.
\cite{Ngo:2015} presented the results 
on searching for stellar companions 
around 50 out of the 51 selected systems 
from \cite{Knutson:2014} study.
They corrected for survey incompleteness
and reported a stellar companion fraction of $48 \pm 9$\%.
Combining the results of both studies,
\cite{Ngo:2015} estimated that $72 \pm 16$\% 
of hot Jupiters are part of multi-planet and/or multi-star systems.

In this work, we report the discovery of \hatcurbc{}, 
the first multi-planet system detected by the HATSouth survey
\citep{bakos:2013:hatsouth}.
The star hosts an inner hot Jupiter
detected via its transits
and an outer cold massive giant planet 
detected via the radial velocity variations of the host star.
The possibility of additional outer planetary companions 
to transiting hot Jupiter has been proposed, 
by e.g. \cite{Rabus:2009} and in fact,
there have been only a few transiting planets
with an outer planetary companion 
for which a full orbit was detected via radial velocity,
such as
HAT-P-13b,c \citep{Bakos:2009},
HAT-P-17b,c \citep{Howard:2012},
Kepler-424b,c \citep{Endl:2014},
WASP-41b,c \citep{Neveu:2016}, 
WASP-47b,c \citep{Hellier:2012,Becker:2015,Neveu:2016},
and WASP-53b,c \citep{Triaud:2017}. 
Among all the systems with a transiting 
hot Jupiter known to have outer companions,
HAT-P-13\,c and WASP-53b,c
are the only massive planetary companions 
with a minimum mass greater than \hatcur{}\,c.
The few detections of companions around
transiting planets is due, to some extent, by the
lack of radial velocity follow-up observations.  
Hot Jupiters in multi-planet systems 
provide a unique opportunity to put observational constraints 
on migration models
and also could be used to probe the
tidal love number of the hot Jupiter
\citep{Buhler:2016,Hardy:2017},
which in turn constrains the planetary interior structure
\citep{Batygin:2009}.
Therefore, monitoring these systems is very interesting
for planet formation and interior structure models. 

The paper is structured as follows:
in Section~\ref{sec:obs}, 
we show the planetary signal detected by the HATSouth network
and present the photometric and spectroscopic follow-up observations
that allowed us to characterize the system.
In Section~\ref{sec:analysis}, 
we derive the stellar parameters
and jointly model the data 
to derive the planetary parameters.
Our results are finally summarized in
Section~\ref{sec:discussion}.


\section{Observations}
\label{sec:obs}

\subsection{Photometry}
\label{sec:photometry}

\subsubsection{Photometric detection}
\label{sec:detection}

\begin{figure}[]
\plotone{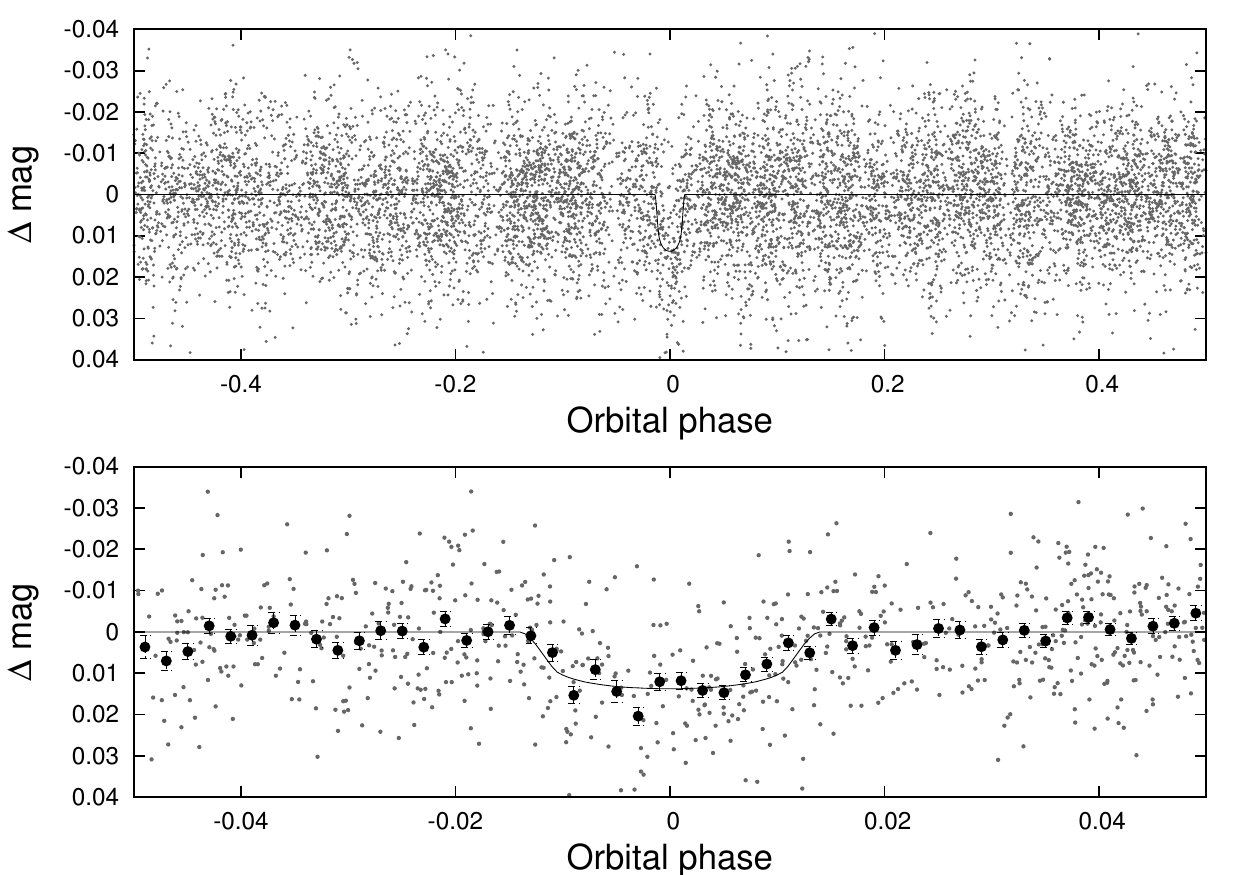}
\caption[]{
        Unbinned instrumental \band{r} \lc{} of \hatcur{} folded with
        the period $P = \hatcurLCPprec$\,days resulting from the global
        fit described in \refsecl{analysis}.  The solid line shows the
        best-fit transit model (see \refsecl{analysis}).  In the lower
        panel we zoom-in on the transit; the dark filled points here
        show the light curve binned in phase using a bin-size of 0.002.\\
\label{fig:hatsouth}}
\end{figure}

The star \hatcur{} (Table~\ref{tab:stellar}) was observed by HATSouth
instruments between UT 2010 January 19 and UT 2010 August 10 using the
HS-1, HS-3, and HS-5 units at Las Campanas Observatory (LCO) in Chile, the H.E.S.S.~site in
Namibia, and Siding Springs Observatory (SSO) in Australia, respectively. A total of 3113, 4690 and
658 images of \hatcur{} were obtained with the HS-1, HS-3 and HS-5 telescopes,
respectively. The observations were obtained through a Sloan $r$
filter with an exposure time of 240\,s. 
The data were reduced to
trend-filtered light curves using the aperture photometry pipeline
described by \citet{penev:2013:hats1} and making use of External
Parameter Decorrelation \citep[EPD;][]{bakos:2010:hat11} and the Trend
Filtering Algorithm \citep[TFA;][]{kovacs:2005:TFA} to remove
systematic variations.  
We searched for transits using the Box Least
Squares \citep[BLS;][]{kovacs:2002:BLS} fitting algorithm, 
and detected a $P=\hatcurLCPshort{}$\,day periodic transit signal 
in the light curve of \hatcur{} 
(Figure~\ref{fig:hatsouth}; the data are available in
Table~\ref{tab:phfu}). 
The per point RMS of the
residual combined filtered HATSouth light curve (after subtracting the
best-fit model transit) is 0.012\,mag, which is typical for a star of
this magnitude.

\subsubsection{Photometric follow-up}
\label{sec:phfu}

\begin{figure}[!ht]
\plotone{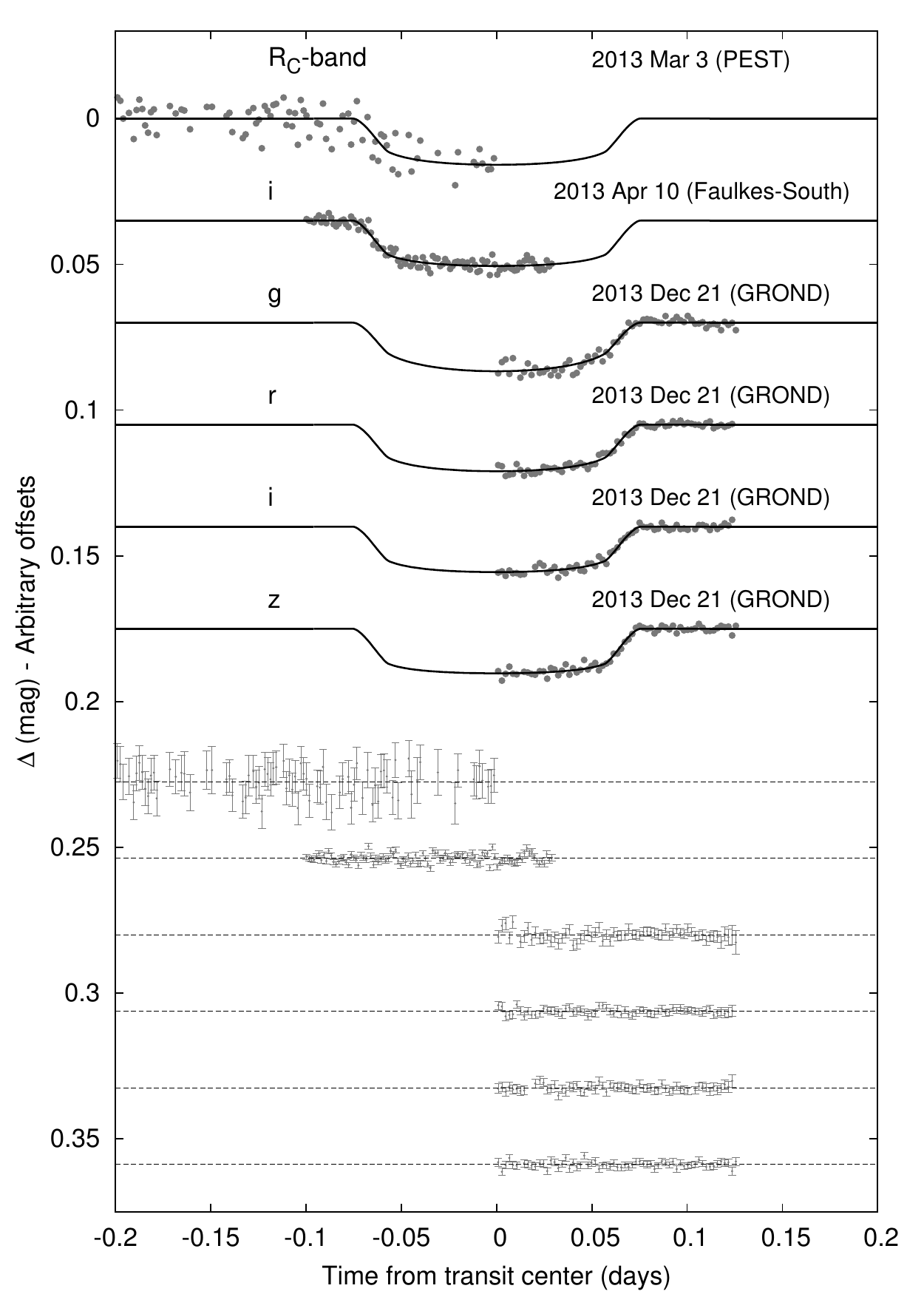}
\caption{
        Unbinned follow-up transit light curve of \hatcur{}. The facilities and filters used, and the dates of each event are listed. Our best fit is
        shown by the solid lines. The residuals from the best-fit
        model are shown below in the same order.\\
\label{fig:lc}} \end{figure}

In order to confirm that the transit signals 
detected in the discovery light curve
are due to a transiting planet,
we obtained photometric follow-up observations 
of three transit events.
These light curves allow us to
refine the ephemeris of the system
and to determine precise parameters of the system.
All the photometric data are provided in Table~\ref{tab:phfu}
and the follow-up light curves are shown in Figure~\ref{fig:lc}
along with the best fit model and residuals.

An ingress was observed with the 
0.3\,m Perth Exoplanet Telescope (PEST)
on 3 March 2013,
using the $R_{C}$ filter.
The photometric precision of the light curve 
was 5.0\,mmag with a cadence of 130\,s.
Another ingress was observed on 
10 April 2013 using the
Faulkes Telescope South (FTS),
which is a fully automated telescope
operated as part of the 
Las Cumbres Observatory Global Telescope
\citep[LCOGT;][]{brown:2013:lcogt}.
The transit was observed in the $i$-band filter
achieving a photometric precision of 1.6\,mmag
with a cadence of 113\,s.
An egress was obtained on 21 December 2013
with the multiband imager GROND \citep{greiner:2008},
mounted on the 2.2\,m telescope in La Silla Observatory,
using four different filters
($g$, $r$, $i$, $z$).
The light curve had a precision of 
1.7\,mmag in the $g$ band,
1.0\,mmag in $r$,
1.1\,mmag in $i$,
and 
1.1\,mmag in $z$,
with a cadence of 168\,s.
The details of the data reduction for these facilities are described in 
\cite{penev:2013:hats1}, \cite{mohlerfischer:2013:hats2} and 
\cite{zhou:2014:mebs}.

\ifthenelse{\boolean{emulateapj}}{
        \begin{deluxetable*}{lrrrrr} }{
        \begin{deluxetable}{lrrrrr} 
    }
        \tablewidth{0pc}
        \tablecaption{Differential photometry of
        \hatcur\label{tab:phfu}} \tablehead{ \colhead{BJD} &
        \colhead{Mag\tablenotemark{a}} &
        \colhead{\ensuremath{\sigma_{\rm Mag}}} &
        \colhead{Mag(orig)\tablenotemark{b}} & \colhead{Filter} &
        \colhead{Instrument} \\ \colhead{\hbox{~~~~(2\,400\,000$+$)~~~~}}
        & \colhead{} & \colhead{} & \colhead{} & \colhead{} &
        \colhead{} } \startdata   
  $ 55372.26299 $ & $ -0.01448 $ & $ 0.00725 $ & $ \cdots $ & $ r$ & HS/G563.1\\
  $ 55274.77568 $ & $ 0.01224 $ & $ 0.00650 $ & $ \cdots $ & $ r$ & HS/G563.1\\
  $ 55296.44071 $ & $ 0.01384 $ & $ 0.00668 $ & $ \cdots $ & $ r$ & HS/G563.1\\
  $ 55274.77891 $ & $ -0.01225 $ & $ 0.00628 $ & $ \cdots $ & $ r$ & HS/G563.1\\
  $ 55296.44428 $ & $ -0.00169 $ & $ 0.00659 $ & $ \cdots $ & $ r$ & HS/G563.1\\
  $ 55274.78240 $ & $ -0.01307 $ & $ 0.00627 $ & $ \cdots $ & $ r$ & HS/G563.1\\
  $ 55296.44754 $ & $ -0.00042 $ & $ 0.00652 $ & $ \cdots $ & $ r$ & HS/G563.1\\
  $ 55274.78561 $ & $ 0.00435 $ & $ 0.00643 $ & $ \cdots $ & $ r$ & HS/G563.1\\
  $ 55296.45080 $ & $ -0.00521 $ & $ 0.00660 $ & $ \cdots $ & $ r$ & HS/G563.1\\
  $ 55372.27744 $ & $ 0.00356 $ & $ 0.00771 $ & $ \cdots $ & $ r$ & HS/G563.1\\
        [-1.5ex]
\enddata \tablenotetext{a}{
     The out-of-transit level has been subtracted. For the HATSouth
     light curve (rows with ``HS'' in the Instrument column), these
     magnitudes have been detrended using the EPD and TFA procedures
     prior to fitting a transit model to the light curve. 
     For the follow-up light curves (rows with an Instrument
     other than ``HS'') these magnitudes have been detrended with the
     EPD procedure, carried out simultaneously with the transit fit.
}
\tablenotetext{b}{
        Raw magnitude values without application of the EPD
        procedure.  This is only reported for the follow-up light
        curve.
}
\tablecomments{
        This table is available in a machine-readable form in the
        online journal.  A portion is shown here for guidance
        regarding its form and content. The data are also available on
        the HATSouth website at \url{http://www.hatsouth.org}.
} \ifthenelse{\boolean{emulateapj}}{ \end{deluxetable*} }{ \end{deluxetable} }

\subsection{Spectroscopic Observations}
\label{sec:hispec}

\begin{figure*} [ht]
\plottwo{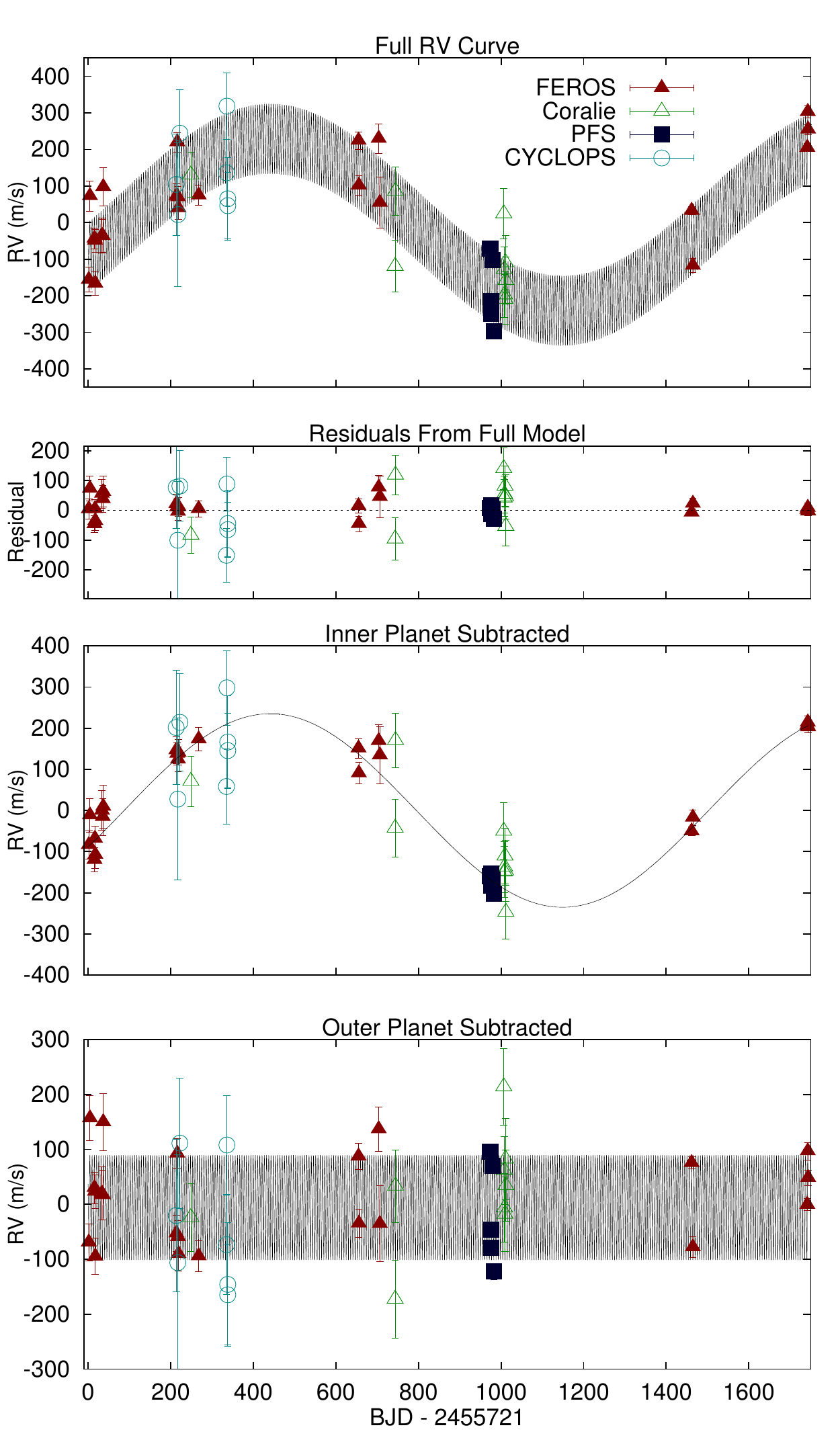}{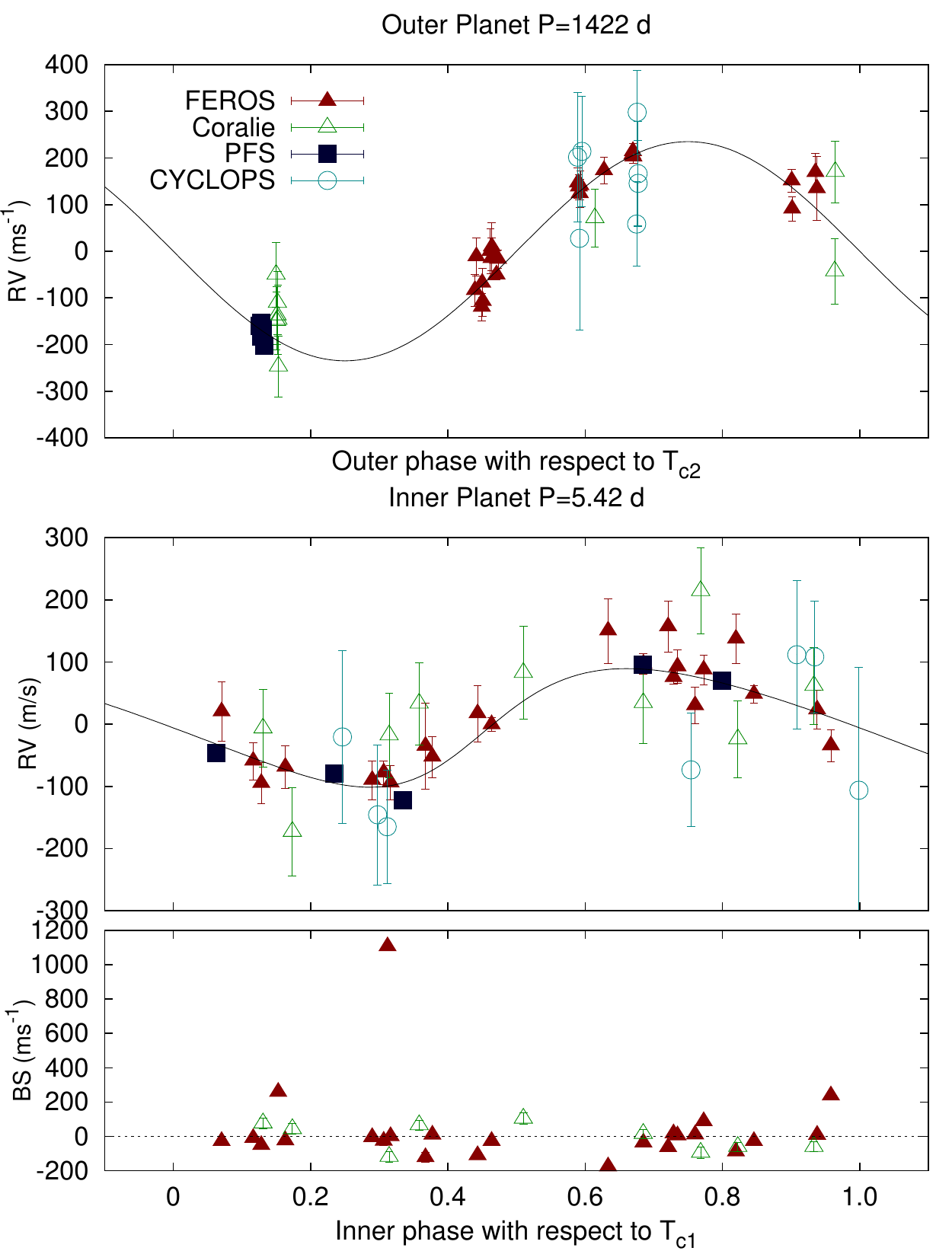}
\caption{
    {\em Top left:} High-precision RV measurements from
    MPG~2.2\,m/FEROS, Euler~1.2\,m/Coralie, Magellan~6.5\,m/PFS, and AAT-3.9m/CYCLOPS, together with our best-fit two-planet orbit model, plotted as a function of time.  The center-of-mass
    velocity has been subtracted. The error bars
    include the jitter which is varied independently for each instrument in the fit. {\em Left, second panel:} RV $O\!-\!C$ residuals from the two planet model, plotted as a function of time. {\em Left, third panel:} RV residuals after subtracting only the model variation due to the inner planet, plotted as a function of time. {\em Left, bottom panel:} RV residuals after subtracting only the model variation due to the outer planet, plotted as a function of time. {\em Top right:} RV residuals after subtracting only the model variation due to the inner planet, plotted as a function of phase of the outer planet. Here zero phase corresponds to the time of inferior conjunction for the outer planet. {\em Right, second panel:} RV residuals after subtracting only the model variation due to the outer planet, plotted as a function of phase of the inner planet. {\em Right, bottom panel:} Spectral line bisector spans (BSs) plotted as a function of phase of the inner planet.   Note the different vertical scales of all of the
    panels.\\
\label{fig:rvbis}}
\end{figure*}

\hatcur{} was spectroscopically observed 
between April 2011 and March 2016
to confirm the planetary nature of the transit signals
and to estimate the mass and therefore the density of the planet.
Furthermore, the long radial velocity (RV) 
monitoring of the star allowed us to 
detect an outer companion with a longer orbital period
than the transiting planet. 
We present the RV used to characterize the system
in Figure~\ref{fig:rvbis} and 
provide the data
in Table~\ref{tab:rvs}.

\subsubsection{Reconnaissance Spectroscopy}

Reconnaissance low-resolution spectroscopic 
follow-up observations are 
important to rule out 
various false positive scenarios, 
such as a primary giant star, 
or large RV variations 
indicating that the transiting object is itself a star.
Reconnaissance spectroscopic observations were carried out 
with the WiFeS spectrograph \citep{dopita:2007}
on the ANU~2.3\,m telescope.
We obtained a single $R = 3000$ spectrum
to estimate the stellar atmospheric parameters
$\teffstar$, $\feh$, and $\vsini$
and were used to confirm that the star is a dwarf.
We also obtained 7 spectra with a higher resolution 
($R=7000$) to look for possible large RV variations
at the $\sim 2$ \kms\ level.
The spectra were extracted and reduced following 
\cite{bayliss:2013:hats3}.
Another reconnaissance spectrum was observed 
with the FIES spectrograph 
at the Nordic Optical Telescope \citep{Telting:2014},
where it was reduced following \cite{buchhave:2010}.
We did not find large RV variations
and thus ruled out the possibility 
that this system might be an eclipsing binary displaying
a large radial velocity amplitude.
We therefore proceeded with acquiring 
high precision RV observations
to characterize the system.

\subsubsection{High Precision Radial Velocities}

We carried out an intensive RV follow-up campaign
to measure, with high precision,
the semi-amplitude of the RV variations
due to the transiting planet. 
The RV observations showed variations in phase 
with the transit ephemeris of the interior planet.
They, additionally, showed evidence
for a large amplitude sinusoidal variation 
with a period of $\sim 1400$\,days.
We next describe the observations 
and the data reduction
of all the spectrographs used in this analysis.

We obtained
9 spectra with CORALIE \citep{queloz:2001},
an echelle fibre-fed spectrograph mounted on the Euler~1.2\,m
in La Silla Observatory.
We also obtained 
5 spectra with the Planet Finder Spectrograph
\citep[PFS,][]{crane:2010} on the Magellan Clay~6.5\,m in Las Campanas Observatory
and 
7 spectra with CYCLOPS on the 
3.9\,m Anglo-Australian Telescope.
Most of the spectra used in this analysis,
most importantly for the discovery of the second outer companion,
were obtained with 
FEROS on the MPG~2.2\,m \citep{kaufer:1998}
in La Silla Observatory.
We observed a total of 24 spectra with FEROS,
which is a high resolution echelle spectrograph \citep{kaufer:1998}.
All the spectra acquired with 
FEROS and CORALIE
were reduced, extracted, and analysed
using the CERES pipeline 
\citep{Brahm:2017:ceres}.
For the data reduction of PFS spectra,
we obtained a template spectrum by using the $0\farcs5$ slit, 
which was then used as reference for computing the radial velocities 
at different epochs by obtaining spectra with an I2-cell. 
The spectra that were acquired with the I2-cell were processed as described in \citet{butler:1996}.
Details on the data reduction and analysis are
described in previous HATSouth discovery papers, e.g.
\cite{jordan:2014:hats4,zhou:2014:hats5,hartman:2015:hats6}.
For details of the data reduction of CYCLOPS spectra,
see \cite{penev:2013:hats1}.

\begin{figure}[t!]
\plotone{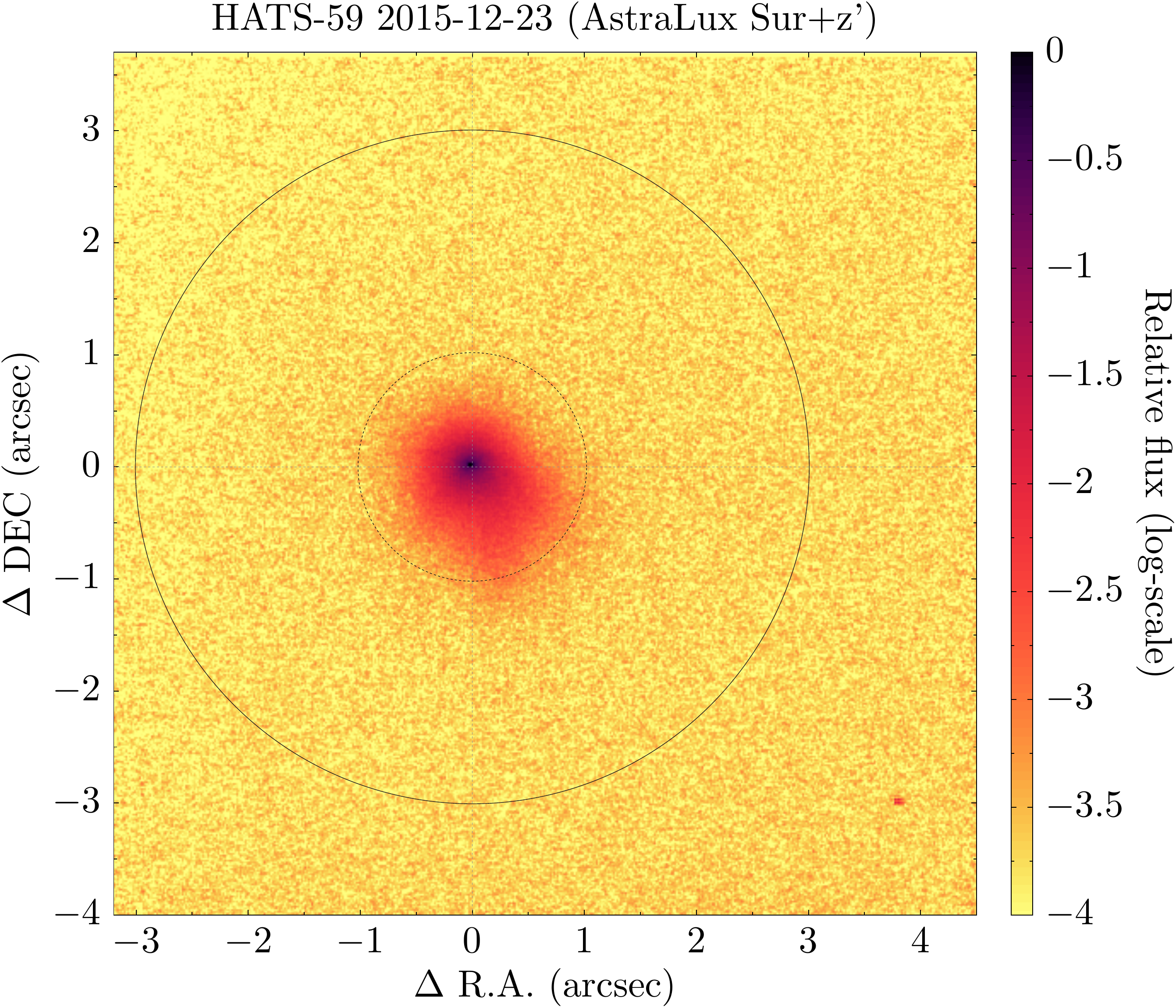}
\caption[]{Astralux Sur lucky image of \hatcur{} using $z'$. Circles of 1$''$ and 3 $''$ radii are shown. No neighboring companion is detected within 2$''$.
\label{fig:image}}
\end{figure}

\begin{figure}[t!]
\plotone{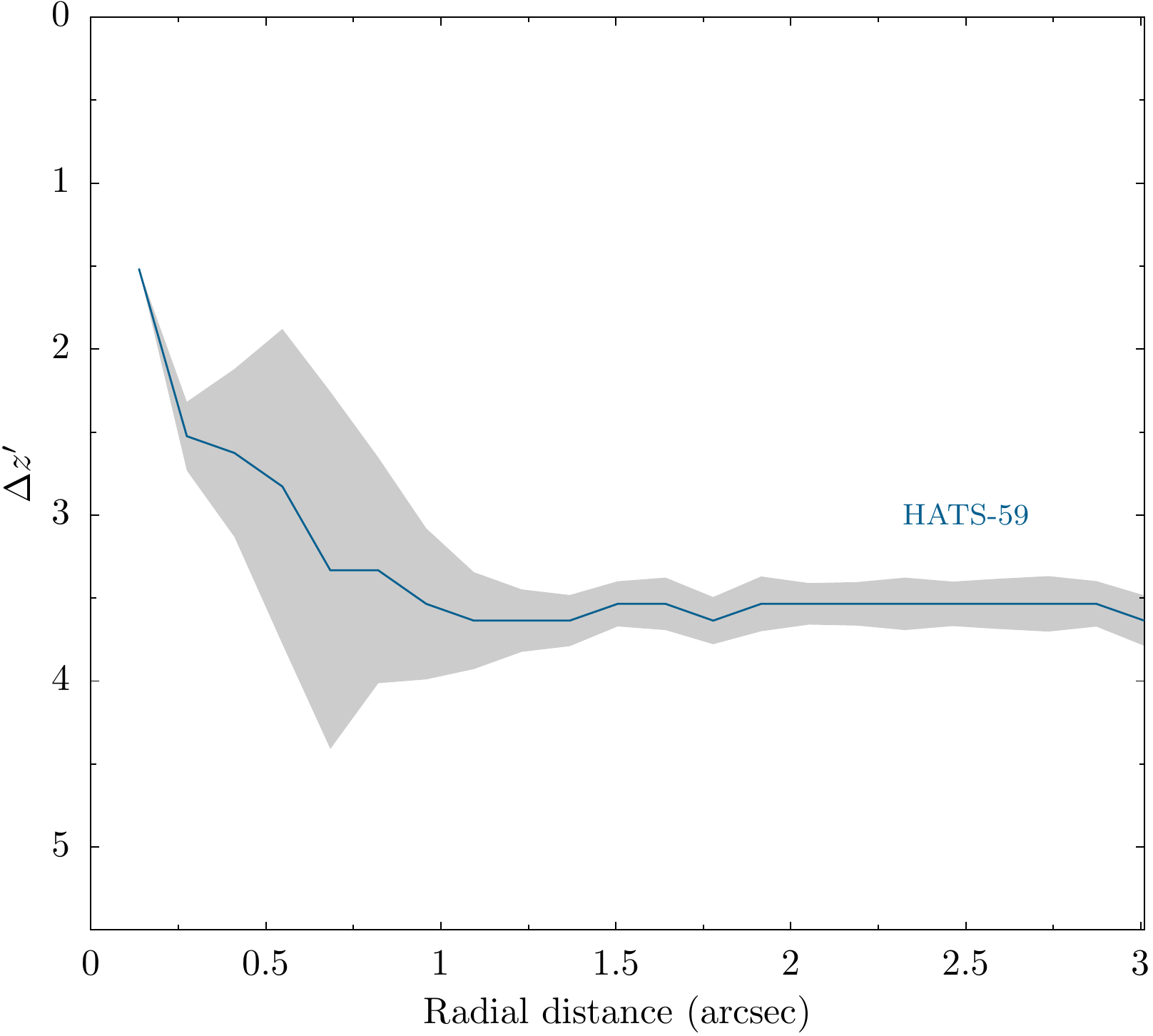}
\caption[]{Contrast curve for of \hatcur{} using the Astralux Sur $z'$ observations. Gray bands show the uncertainty given by the scatter in the contrast in the azimuthal direction at a given radius.
\label{fig:contrastcurve}}
\end{figure}

\subsection{Lucky Imaging}
\label{sec:lucky}

High spatial resolution imaging were obtained 
as part of the follow-up campaign 
using the Astralux Sur camera \citep{Hippler:2009}
on the New Technology Telescope (NTT),
at La Silla Observatory in Chile. 
These observations are useful to 
identify close stellar companions 
that could affect the transit depth.
Lucky imaging observations were obtained on
December 23, 2015 using the SDSS $z'$ filter
and reduced following \cite{Espinoza:2016}
but we used instead the plate scale derived in 
\cite{Janson:2017} of 15.2 mas/pixel,
which is a better estimate than the one 
estimated in our previous work.
Figure~\ref{fig:image} shows the final reduced image
and Figure~\ref{fig:contrastcurve}
shows the contrast curve,
where no resolved companion is detected within 2$''$.

%

\ifthenelse{\boolean{emulateapj}}{
    \begin{deluxetable*}{lrrrrrr}
}{
    \begin{deluxetable}{lrrrrrr}
}
\tablewidth{0pc}
\tablecaption{
    Relative radial velocities and bisector span measurements of
    \hatcur{}.
    \label{tab:rvs}
}
\tablehead{
    \colhead{BJD} & 
    \colhead{RV\tablenotemark{a}} & 
    \colhead{\ensuremath{\sigma_{\rm RV}}\tablenotemark{b}} & 
    \colhead{BS} & 
    \colhead{\ensuremath{\sigma_{\rm BS}}} & 
        \colhead{Phase} &
        \colhead{Instrument}\\
    \colhead{\hbox{(2\,450\,000$+$)}} & 
    \colhead{(\ms)} & 
    \colhead{(\ms)} &
    \colhead{(\ms)} &
    \colhead{} &
        \colhead{} &
        \colhead{}
}
\startdata
  $ 5722.48192 $ & $ -156.41 $ & $ 34.00 $ & $ -25.0 $ & $ 16.0 $ & $ 0.163 $ & FEROS \\
  $ 5725.50346 $ & $ 72.59 $ & $ 41.00 $ & $ -64.0 $ & $ 19.0 $ & $ 0.721 $ & FEROS \\
  $ 5736.54834 $ & $ -43.41 $ & $ 29.00 $ & $ 10.0 $ & $ 14.0 $ & $ 0.760 $ & FEROS \\
  $ 5737.51083 $ & $ -49.41 $ & $ 31.00 $ & $ 7.0 $ & $ 15.0 $ & $ 0.938 $ & FEROS \\
  $ 5738.54029 $ & $ -166.41 $ & $ 33.00 $ & $ -52.0 $ & $ 16.0 $ & $ 0.128 $ & FEROS \\
  $ 5754.47565 $ & $ -35.41 $ & $ 48.00 $ & $ -29.0 $ & $ 21.0 $ & $ 0.070 $ & FEROS \\
  $ 5756.49554 $ & $ -36.41 $ & $ 45.00 $ & $ -111.0 $ & $ 20.0 $ & $ 0.443 $ & FEROS \\
  $ 5757.52534 $ & $ 97.59 $ & $ 52.00 $ & $ -177.0 $ & $ 21.0 $ & $ 0.633 $ & FEROS \\
  $ 5934.15946 $ & $ 104.14 $ & $ 109.00 $ & \nodata & \nodata & $ 0.246 $ & CYCLOPS \\
  $ 5934.86914 $ & $ 72.59 $ & $ 33.00 $ & $ 8.0 $ & $ 16.0 $ & $ 0.377 $ & FEROS \\
  $ 5936.80355 $ & $ 219.59 $ & $ 27.00 $ & $ 4.0 $ & $ 13.0 $ & $ 0.735 $ & FEROS \\
  $ 5938.23445 $ & $ 22.14 $ & $ 177.00 $ & \nodata & \nodata & $ 0.999 $ & CYCLOPS \\
  $ 5938.87128 $ & $ 69.59 $ & $ 30.00 $ & $ -11.0 $ & $ 14.0 $ & $ 0.116 $ & FEROS \\
  $ 5939.81042 $ & $ 39.59 $ & $ 31.00 $ & $ -6.0 $ & $ 15.0 $ & $ 0.290 $ & FEROS \\
  $ 5943.16020 $ & $ 244.14 $ & $ 82.00 $ & \nodata & \nodata & $ 0.908 $ & CYCLOPS \\
  $ 5969.77597 $ & $ 130.39 $ & $ 36.00 $ & $ -61.0 $ & $ 26.0 $ & $ 0.822 $ & Coralie \\
  $ 5988.70024 $ & $ 74.59 $ & $ 28.00 $ & $ 0.0 $ & $ 14.0 $ & $ 0.317 $ & FEROS \\
  $ 6056.06292 $ & $ 136.14 $ & $ 29.00 $ & \nodata & \nodata & $ 0.754 $ & CYCLOPS \\
  $ 6057.03928 $ & $ 318.14 $ & $ 27.00 $ & \nodata & \nodata & $ 0.934 $ & CYCLOPS \\
  $ 6059.00663 $ & $ 65.14 $ & $ 73.00 $ & \nodata & \nodata & $ 0.298 $ & CYCLOPS \\
  $ 6059.08191 $ & $ 46.14 $ & $ 30.00 $ & \nodata & \nodata & $ 0.312 $ & CYCLOPS \\
  $ 6375.71072 $ & $ 224.59 $ & $ 24.00 $ & $ 87.0 $ & $ 12.0 $ & $ 0.773 $ & FEROS \\
  $ 6376.71477 $ & $ 101.59 $ & $ 26.00 $ & $ 237.0 $ & $ 13.0 $ & $ 0.958 $ & FEROS \\
  $ 6377.76897 $ & \nodata & \nodata & $ 258.0 $ & $ 12.0 $ & $ 0.153 $ & FEROS \\
  $ 6378.63214 $ & \nodata & \nodata & $ 1107.0 $ & $ 13.0 $ & $ 0.312 $ & FEROS \\
  $ 6424.70951 $ & $ 229.59 $ & $ 40.00 $ & $ -90.0 $ & $ 18.0 $ & $ 0.820 $ & FEROS \\
  $ 6427.67642 $ & $ 54.59 $ & $ 69.00 $ & $ -123.0 $ & $ 30.0 $ & $ 0.367 $ & FEROS \\
  $ 6464.53773 $ & $ -119.61 $ & $ 50.00 $ & $ 43.0 $ & $ 32.0 $ & $ 0.173 $ & Coralie \\
  $ 6465.53888 $ & $ 85.39 $ & $ 43.00 $ & $ 65.0 $ & $ 29.0 $ & $ 0.358 $ & Coralie \\
  $ 6694.77700 $ & $ -71.73 $ & $ 6.28 $ & \nodata & \nodata & $ 0.684 $ & PFS \\
  $ 6696.82969 $ & $ -214.95 $ & $ 4.90 $ & \nodata & \nodata & $ 0.063 $ & PFS \\
  $ 6697.75879 $ & $ -248.73 $ & $ 4.86 $ & \nodata & \nodata & $ 0.234 $ & PFS \\
  $ 6700.82214 $ & $ -101.76 $ & $ 5.72 $ & \nodata & \nodata & $ 0.800 $ & PFS \\
  $ 6703.71677 $ & $ -296.25 $ & $ 8.71 $ & \nodata & \nodata & $ 0.334 $ & PFS \\
  $ 6727.73071 $ & $ 24.39 $ & $ 48.00 $ & $ -95.0 $ & $ 32.0 $ & $ 0.768 $ & Coralie \\
  $ 6728.62473 $ & $ -128.61 $ & $ 36.00 $ & $ -61.0 $ & $ 29.0 $ & $ 0.933 $ & Coralie \\
  $ 6729.69322 $ & $ -197.61 $ & $ 37.00 $ & $ 76.0 $ & $ 29.0 $ & $ 0.131 $ & Coralie \\
  $ 6730.68884 $ & $ -209.61 $ & $ 46.00 $ & $ -119.0 $ & $ 32.0 $ & $ 0.314 $ & Coralie \\
  $ 6731.74803 $ & $ -109.61 $ & $ 55.00 $ & $ 104.0 $ & $ 32.0 $ & $ 0.510 $ & Coralie \\
  $ 6732.69422 $ & $ -158.61 $ & $ 41.00 $ & $ 15.0 $ & $ 26.0 $ & $ 0.685 $ & Coralie \\
  $ 7182.46643 $ & $ 32.59 $ & $ 11.00 $ & $ 16.0 $ & $ 15.0 $ & $ 0.729 $ & FEROS \\
  $ 7185.59484 $ & $ -117.41 $ & $ 19.00 $ & $ -27.0 $ & $ 25.0 $ & $ 0.306 $ & FEROS \\
  $ 7462.66518 $ & $ 204.59 $ & $ 11.00 $ & $ -28.0 $ & $ 16.0 $ & $ 0.464 $ & FEROS \\
  $ 7463.86306 $ & $ 302.59 $ & $ 16.00 $ & $ -37.0 $ & $ 21.0 $ & $ 0.685 $ & FEROS \\
  $ 7464.73538 $ & $ 254.59 $ & $ 14.00 $ & $ -29.0 $ & $ 19.0 $ & $ 0.846 $ & FEROS \\
    [-1.5ex]
\enddata
\tablenotetext{a}{
        Relative RVs, with $\gamma_{RV}$ subtracted.
}
\tablenotetext{b}{
        Internal errors excluding the component of
        astrophysical/instrumental jitter considered in
        \refsecl{analysis}.
}
\ifthenelse{\boolean{emulateapj}}{
    \end{deluxetable*}
}{
    \end{deluxetable}
}

\section{Analysis}
\label{sec:analysis}

\begin{figure}[]
\plotone{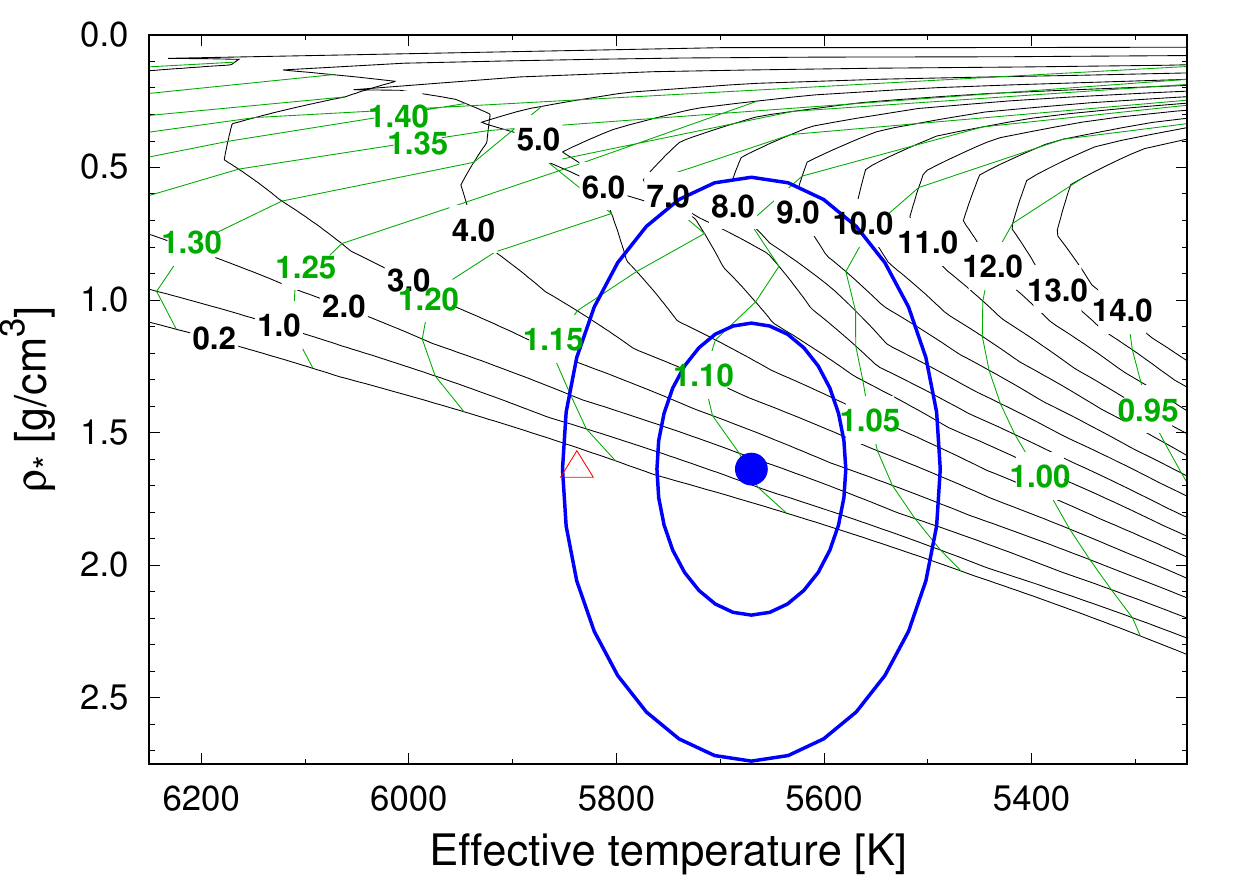}
\caption[]{
    Model isochrones (black solid lines) from \cite{\hatcurisocite} 
    for the measured metallicity of \hatcur. 
    The age of each isochrone in Gyr is labeled in black font. 
    We also show evolutionary tracks for stars of fixed mass (dashed green lines)
    with the mass of each tracked labeled in solar mass units in green font. 
    The adopted values of $\teffstar$ and \rhostar\ are shown 
    using the filled blue circle together with
    their 1$\sigma$ and 2$\sigma$ confidence ellipsoids (blue lines).  The initial
    values of \teffstar\ and \rhostar\ from the first ZASPE and \lc\
    analysis are represented with the red open triangle.
\label{fig:iso}}
\end{figure}

\subsection{Properties of the Parent Star}

It is important to characterize the host star
in order to measure precise planetary parameters.
We used ZASPE \citep{brahm:2017:zaspe}
to get an initial estimate of the atmospheric parameters
(\teffstar\, \feh\, $\vsini$, and \loggstar).
The parameters 
were determined using the FEROS spectra,
which were co-added to obtain a high 
signal-to-noise ratio spectrum.
ZASPE determines the stellar parameters
via least-squares minimization
against a grid of synthetic spectra 
in the spectral regions 
most sensitive to changes in the parameters
(5000 \AA\ and 6000 \AA).

We then followed \cite{sozzetti:2007}
to determine the fundamental stellar parameters
($\mstar$, $\rstar$, $\lstar$, age, etc.).
In particular, we used the stellar density \rhostar\,
determined from the photometric light curve,
combined with the \teffstar\ and \feh\ measurements,
to characterize the host star.
The parameters were obtained by combining 
the spectroscopic and photometric parameters with the 
Yonsei-Yale stellar evolution models \citep[Y$^{2}$;][]{yi:2001}.
This provided a revised
estimate of \loggstar,
which was fixed in a second iteration of ZASPE
that returned the final values of the stellar parameters.

We find that the star \hatcur{} has a mass of
\hatcurISOm\,\msun, a radius of \hatcurISOr\,\rsun, and is at a
reddening-corrected distance of \hatcurXdistred\,pc.
The distance estimated using isochrone fitting 
is in agreement with the distance estimated using Gaia data.
Figure~\ref{fig:iso} shows the location
of the star on the \teffstar-\rhostar\ diagram
and the final adopted stellar parameters are listed in
Table~\ref{tab:stellar}.

\subsection{Excluding Blend Scenarios}

In order to rule out the possibility that \hatcur{} is a blended
stellar eclipsing binary system, we carried out a blend analysis of
the photometric data following \citet{hartman:2012:hat39hat41}. We
find that although blended stellar eclipsing binary models can be
found which fit the available photometric data, these models would
produce obviously composite spectroscopic cross-correlation functions
(CCFs) that are inconsistent with the observed CCFs. For example, in
all cases the spectral line bisector spans (BSs) computed from the
simulated CCFs have scatter in excess of 10\,\kms, whereas, excluding
one outlier, the scatter of the measured FEROS BSs is $\sim
100$\,\ms. Similarly the RVs of the simulated CCFs are in excess of
500\,\ms, whereas the observed FEROS RVs have a scatter of
130\,\ms\ (dominated by the planetary signals). We conclude that
\hatcur{} is not a blended stellar eclipsing binary, and is instead a
transiting planet system.

\subsection{Global Modeling of the Data}

To measure the 
orbital and physical properties of the planets,
we modeled all the photometric data
(the HATSouth and follow-up photometric data)
and the high-precision RV measurements
following \cite{pal:2008:hat7} \cite{bakos:2010:hat11}
and \cite{hartman:2012:hat39hat41}.

All the photometric light curves
were modeled using the \cite{mandel:2002} transit models
with fixed quadratic limb darkening coefficients
taken from \cite{claret:2004}.
For the HATSouth discovery photometric light curves,
we also considered a dilution factor for the transit depth
that accounts for possible blends from neighboring stars 
and possible over-correction 
introduced by the trend filtering algorithm
\citep[TFA; removes trends shared with other stars;][]{bakos:2010:hat11,kovacs:2005:TFA}.
As for the photometric follow-up light curves,
the systematic trends were corrected by including 
a quadratic trend to the transit model.
We also added a linear trend, with up to three parameters,
to reconstruct the shape of the PSF.
This trend compensates for changes in the PSF 
during the observations, which could be due to 
poor guiding, non-photometric conditions, or 
changes in the seeing during the transit observations.

We fit the RVs,
taken with different spectrographs,
with a Keplerian orbit
allowing the zero-point and the RV jitter, 
for each instrument, 
to vary independently in the fit.
Our RVs support the existence of a second planet 
on top of the transiting one, 
and therefore models with two planets were considered in the modelling.
We considered four different scenarios 
where one or both of the planets 
had a fixed circular orbit, 
or was allowed to have non-zero eccentricity.  
To choose between the different scenarios, 
we estimated the Bayesian evidence for each model 
following \citet{weinberg:2013}, 
and then adopted the model with the highest evidence, 
which we find to be a model in which 
the interior transiting planet has a non-zero eccentricity,
while the exterior planet has a circular orbit.
The evidence for this model is a modest factor of 2.4 times 
greater than the evidence 
for the model in which both planets are assumed to have circular orbits, 
7 times greater than the model in which 
the interior planet is circular 
and the exterior planet has an eccentric orbit, 
and 19 times greater than the model in which 
both planets have non-zero eccentricities. 

This analysis makes use of a 
differential evolution Markov Chain Monte Carlo procedure 
\citep[DEMCMC;][]{terbraak:2006} 
to estimate the posterior parameter distributions, 
which we use to determine 
the median parameter values and their 1$\sigma$ uncertainties. 
We find that the transiting planet \hatcurb{} has a
mass of \hatcurPPmlong\,\mjup, 
a radius of \hatcurPPrlong\,\rjup, 
and a non-zero eccentricity of $e = \hatcurRVeccen$. 
For the second planet, 
which we dub \hatcurc{}, 
we find that is well fit by a circular Keplerian orbit 
with $P = \hatcurcLCP$\,days, 
$K = \hatcurcRVK{}$\,\ms, 
implying a minimum mass for the companion of $\msini = \hatcurcPPm{}$\,\mjup,
where $i$ is the orbital inclination of \hatcurc{}.

\ifthenelse{\boolean{emulateapj}}{
  \begin{deluxetable*}{lcr}
}{
  \begin{deluxetable}{lcr}
}
\tablewidth{0pc}
\tabletypesize{\scriptsize}
\tablecaption{
    Stellar Parameters for \hatcur{} 
    \label{tab:stellar}
}
\tablehead{
    \multicolumn{1}{c}{~~~~~~~~Parameter~~~~~~~~} &
    \multicolumn{1}{c}{Value}                     &
    \multicolumn{1}{c}{Source}    
}
\startdata
\noalign{\vskip -3pt}
\sidehead{Identifying Information}
~~~~R.A.~(h:m:s)                      &  \hatcurCCra{} & 2MASS\\
~~~~Dec.~(d:m:s)                      &  \hatcurCCdec{} & 2MASS\\
~~~~R.A.p.m.~(mas/yr)                 &  $-24.16 \pm 0.047$ & Gaia DR2 \\
~~~~Dec.p.m.~(mas/yr)                 &  $ 0.92 \pm 0.03 $ & Gaia DR2 \\
~~~~Parallax~(mas)                    &  $ 1.52 \pm 0.03 $ & Gaia DR2 \\
~~~~GSC ID                            &  \hatcurCCgsc{} & GSC\\
~~~~2MASS ID                          &  \hatcurCCtwomass{} & 2MASS\\
\sidehead{Spectroscopic properties}
~~~~$\teffstar$ (K)\dotfill         &  \hatcurSMEteff{} & ZASPE \tablenotemark{a}\\
~~~~$\feh$\dotfill                  &  \hatcurSMEzfeh{} & ZASPE                 \\
~~~~$\vsini$ (\kms)\dotfill         &  \hatcurSMEvsin{} & ZASPE                 \\
~~~~$\gamma_{\rm RV}$ (\ms)\dotfill&  \hatcurRVgammaabs{} & FEROS                  \\
\sidehead{Photometric properties}
~~~~$B$ (mag)\dotfill               &  \hatcurCCtassmB{} & APASS                \\
~~~~$V$ (mag)\dotfill               &  \hatcurCCtassmv{} & APASS               \\
~~~~$g$ (mag)\dotfill               &  \hatcurCCtassmg{} & APASS                \\
~~~~$r$ (mag)\dotfill               &  \hatcurCCtassmr{} & APASS                \\
~~~~$i$ (mag)\dotfill               &  \hatcurCCtassmi{} & APASS                \\
~~~~$J$ (mag)\dotfill               &  \hatcurCCtwomassJmag{} & 2MASS           \\
~~~~$H$ (mag)\dotfill               &  \hatcurCCtwomassHmag{} & 2MASS           \\
~~~~$K_s$ (mag)\dotfill             &  \hatcurCCtwomassKmag{} & 2MASS           \\
~~~~$G$ (mag)\dotfill               &  $13.785 $  & Gaia DR2        \\
\sidehead{Derived properties}
~~~~$\mstar$ ($\msun$)\dotfill      &  \hatcurISOmlong{} & Y$^{2}$+\hatcurlumind{}+ZASPE \tablenotemark{b}\\
~~~~$\rstar$ ($\rsun$)\dotfill      &  \hatcurISOrlong{} & Y$^{2}$+\hatcurlumind{}+ZASPE         \\
~~~~$\loggstar$ (cgs)\dotfill       &  \hatcurISOlogg{} & Y$^{2}$+\hatcurlumind{}+ZASPE         \\
~~~~$\rhostar$ (\gcmc) \tablenotemark{c}\dotfill & \hatcurLCrho{} & Light curves \\
~~~~$\rhostar$ (\gcmc) \tablenotemark{c}\dotfill & \hatcurISOrho{} & Y$^{2}$+Light curves+ZASPE \\
~~~~$\lstar$ ($\lsun$)\dotfill      &  \hatcurISOlum{} & Y$^{2}$+\hatcurlumind{}+ZASPE         \\
~~~~$M_V$ (mag)\dotfill             &  \hatcurISOmv{} & Y$^{2}$+\hatcurlumind{}+ZASPE         \\
~~~~$M_K$ (mag,\hatcurjhkfilset{})&  \hatcurISOMK{} & Y$^{2}$+\hatcurlumind{}+ZASPE         \\
~~~~Age (Gyr)\dotfill               &  \hatcurISOage{} & Y$^{2}$+\hatcurlumind{}+ZASPE         \\
~~~~$A_{V}$ (mag) \tablenotemark{d}\dotfill           &  \hatcurXAv{} & Y$^{2}$+\hatcurlumind{}+ZASPE\\
~~~~Distance (pc)\dotfill           &  $ 654 \pm 14 $ & Gaia DR2\\
\enddata
\tablenotetext{a}{
    ZASPE = ``Zonal Atmospherical Stellar Parameter Estimator'' method
    for the analysis of high-resolution spectra \citep{brahm:2017:zaspe}
    applied to the FEROS spectra of \hatcur{}. These parameters rely
    primarily on ZASPE, but have a small dependence also on the
    iterative analysis incorporating the isochrone search and global
    modeling of the data, as described in the text.  }
\tablenotetext{b}{
    Isochrones+\hatcurlumind{}+ZASPE = Based on the Y$^{2}$ isochrones
    \citep{yi:2001},
    the stellar density used as a luminosity indicator, and the ZASPE
    results.
} 
\tablenotetext{c}{We list two values for $\rhostar$. The first value is determined from the global fit to the light curves and RV data, without imposing a constraint that the parameters match the stellar evolution models. The second value results from restricting the posterior distribution to combinations of $\rhostar$+$\teffstar$+$\feh$ that match to a Y$^{2}$ stellar model.}
\tablenotetext{d}{ Total \band{V} extinction to the star determined
  by comparing the catalog broad-band photometry listed in the table
  to the expected magnitudes from the
  Isochrones+\hatcurlumind{}+ZASPE model for the star. We use the
  \citet{cardelli:1989} extinction law.  }
\ifthenelse{\boolean{emulateapj}}{
  \end{deluxetable*}
}{
  \end{deluxetable}
}

\ifthenelse{\boolean{emulateapj}}{
  \begin{deluxetable*}{lrr}
}{
  \begin{deluxetable}{lrr}
}
\tabletypesize{\scriptsize}
\tablecaption{Parameters for the planets \hatcurbc{}.
\label{tab:planetparam}}
\tablehead{
    \multicolumn{1}{c}{~~~~~~~~Parameter~~~~~~~~} &
    \multicolumn{1}{r}{\hatcurb{}} &
    \multicolumn{1}{r}{\hatcurc{}} \\
    \multicolumn{1}{c}{} &
    \multicolumn{1}{r}{Value \tablenotemark{a}} &                
    \multicolumn{1}{r}{Value \tablenotemark{a}}                     
}
\startdata
\noalign{\vskip -3pt}
\sidehead{\Lc{} parameters}
~~~$P$ (days)             \dotfill    & $\hatcurLCP{}$ & $\hatcurcLCP{}$             \\
~~~$T_c$ (${\rm BJD}$)    
      \tablenotemark{b}   \dotfill    & $\hatcurLCT{}$ & $\hatcurcLCT{}$              \\
~~~$T_{14}$ (days)
      \tablenotemark{b}   \dotfill    & $\hatcurLCdur{}$ & $\hatcurcLCdur{}$            \\
~~~$T_{12} = T_{34}$ (days)
      \tablenotemark{b}   \dotfill    & $\hatcurLCingdur{}$ & $\hatcurcLCingdur{}$         \\
~~~$\arstar$              \dotfill    & $\hatcurPPar{}$ & $\hatcurcPPar{}$             \\
~~~$\zrstar$ \tablenotemark{c}              \dotfill    & $\hatcurLCzeta{}$  & $\cdots$      \\
~~~$\rpl/\rstar$          \dotfill    & $\hatcurLCrprstar{}$  & $\cdots$      \\
~~~$b^2$                  \dotfill    & $\hatcurLCbsq{}$  & $\cdots$          \\
~~~$b \equiv a \cos i/\rstar$
                          \dotfill    & $\hatcurLCimp{}$ & $\cdots$          \\
~~~$i$ (deg)              \dotfill    & $\hatcurPPi{}$ & $\cdots$        \\

\sidehead{Limb-darkening coefficients \tablenotemark{d}}
~~~$c_1,g$ (linear term) \dotfill    & $\hatcurLBig{}$    & $\cdots$         \\
~~~$c_2,g$ (quadratic term) \dotfill    & $\hatcurLBiig{}$   & $\cdots$         \\
~~~$c_1,R$              \dotfill    & $\hatcurLBiR{}$    & $\cdots$         \\
~~~$c_2,R$               \dotfill    & $\hatcurLBiiR{}$   & $\cdots$         \\
~~~$c_1,r$              \dotfill    & $\hatcurLBir{}$    & $\cdots$         \\
~~~$c_2,r$               \dotfill    & $\hatcurLBiir{}$   & $\cdots$         \\
~~~$c_1,i$ \dotfill    & $\hatcurLBii{}$   & $\cdots$         \\
~~~$c_2,i$ \dotfill  & $\hatcurLBiii{}$  & $\cdots$         \\
~~~$c_1,z$              \dotfill    & $\hatcurLBiz{}$    & $\cdots$         \\
~~~$c_2,z$               \dotfill    & $\hatcurLBiiz{}$   & $\cdots$         \\

\sidehead{RV parameters}
~~~$K$ (\ms)              \dotfill    & $\hatcurRVK{}$ & $\hatcurcRVK{}$      \\
~~~$e$ \tablenotemark{e}  \dotfill    & $\hatcurRVeccen{}$ & $\hatcurcRVeccentwosiglimecceneccen{}$ \\
~~~$\omega$  \dotfill    & $\hatcurRVomega{}$ & $\cdots$ \\
~~~$\sqrt{e}\cos\omega$  \dotfill    & $\hatcurRVrk{}$ & $\cdots$ \\
~~~$\sqrt{e}\sin\omega$  \dotfill    & $\hatcurRVrh{}$ & $\cdots$ \\
~~~$e\cos\omega$  \dotfill    & $\hatcurRVk{}$ & $\cdots$ \\
~~~$e\sin\omega$  \dotfill    & $\hatcurRVh{}$ & $\cdots$ \\
~~~FEROS RV jitter (\ms) \tablenotemark{f}        \dotfill    & \hatcurRVjittertwosiglimA{} & $\cdots$           \\
~~~Coralie RV jitter (\ms) \tablenotemark{f}        \dotfill    & \hatcurRVjitterB{} & $\cdots$           \\
~~~PFS RV jitter (\ms) \tablenotemark{f}        \dotfill    & \hatcurRVjitterC{} & $\cdots$           \\
~~~CYCLOPS RV jitter (\ms) \tablenotemark{f}        \dotfill    & \hatcurRVjitterD{} & $\cdots$           \\

\sidehead{Planetary parameters}
~~~$\mpl$ ($\mjup$)       \dotfill    & $\hatcurPPmlong{}$ & $\cdots$         \\
~~~$\mpl\sin i$ ($\mjup$)       \dotfill    & $\cdots$ & $\hatcurcPPmlong{}$          \\
~~~$\rpl$ ($\rjup$)       \dotfill    & $\hatcurPPrlong{}$ & $\cdots$         \\
~~~$C(\mpl,\rpl)$
    \tablenotemark{g}     \dotfill    & $\hatcurPPmrcorr{}$ & $\cdots$        \\
~~~$\rhopl$ (\gcmc)       \dotfill    & $\hatcurPPrho{}$ & $\cdots$           \\
~~~$\log g_p$ (cgs)       \dotfill    & $\hatcurPPlogg{}$ & $\cdots$          \\
~~~$a$ (AU)               \dotfill    & $\hatcurPParel{}$ & $\hatcurcPParel{}$         \\
~~~$T_{\rm eq}$ (K) \tablenotemark{h}        \dotfill   & $\hatcurPPteff{}$  & $\hatcurcPPteff{}$        \\
~~~$\Theta$ \tablenotemark{i} \dotfill & $\hatcurPPtheta{}$  & $\cdots$      \\
~~~$\langle F \rangle$ (\ergscmsq) \tablenotemark{i}
                          \dotfill    & $(\hatcurPPfluxavg{}) \times 10^{\hatcurPPfluxavgdim}$ & $(\hatcurcPPfluxavg{}) \times 10^{\hatcurcPPfluxavgdim}$      \\ [-1.5ex]
\enddata
\tablenotetext{a}{
    For each parameter we give the median value and
    68.3\% (1$\sigma$) confidence intervals from the posterior
    distribution. Reported results assume an eccentric orbit for \hatcurb{} and a circular orbit for \hatcurc{}.
}
\tablenotetext{b}{
    Reported times are in Barycentric Julian Date calculated directly
    from UTC, {\em without} correction for leap seconds.
    \ensuremath{T_c}: Reference epoch of mid transit that
    minimizes the correlation with the orbital period. Note that \hatcurc{} has not been observed to transit. We list here the time of mid transit, implied by the orbital solution, in the event that the orbital inclination permits transits.
    \ensuremath{T_{14}}: total transit duration, time
    between first to last contact; 
    \ensuremath{T_{12}=T_{34}}: ingress/egress time, time between first
    and second, or third and fourth contact. For \hatcurc{} \ensuremath{T_{14}} and \ensuremath{T_{12}} are calculated assuming central transits ($i = 90^{\circ}$ orbit) and a Jupiter radius for the planet.
}
\tablenotetext{c}{
    Reciprocal of the half duration of the transit used as a jump
    parameter in our MCMC analysis in place of $\arstar$. It is
    related to $\arstar$ by the expression $\zrstar = \arstar
    (2\pi(1+e\sin \omega))/(P \sqrt{1 - b^{2}}\sqrt{1-e^{2}})$
    \citep{bakos:2010:hat11}.
}
\tablenotetext{d}{
    Values for a quadratic law, adopted from the tabulations by
    \cite{claret:2004} according to the spectroscopic (ZASPE) parameters
    listed in \reftabl{stellar}.
}
\tablenotetext{e}{
    For \hatcurc{} we list the 95\% confidence upper-limit on the eccentricity. All other
    parameters listed are determined assuming a circular orbit for this planet.
}
\tablenotetext{f}{
    Error term, either astrophysical or instrumental in origin, added
    in quadrature to the formal RV errors. This term is varied in the
    fit independently for each instrument assuming a prior that is inversely proportional to the jitter.
}
\tablenotetext{g}{
    Correlation coefficient between the planetary mass \mpl\ and
    radius \rpl\ determined from the parameter posterior distribution
    via $C(\mpl,\rpl) = \langle(\mpl - \langle\mpl\rangle)(\rpl -
    \langle\rpl\rangle)\rangle/(\sigma_{\mpl}\sigma_{\rpl})\rangle$, 
	where $\langle \cdot \rangle$ is the
    expectation value, and $\sigma_x$ is the std.\
    dev.\ of $x$.
}
\tablenotetext{h}{
    Planet equilibrium temperature averaged over the orbit, calculated
    assuming a Bond albedo of zero, and that flux is reradiated from
    the full planet surface.
}
\tablenotetext{i}{
    The Safronov number is given by $\Theta = \frac{1}{2}(V_{\rm
    esc}/V_{\rm orb})^2 = (a/\rpl)(\mpl / \mstar )$
    \citep[see][]{hansen:2007}.
}
\tablenotetext{j}{
    Incoming flux per unit surface area, averaged over the orbit.
}
\ifthenelse{\boolean{emulateapj}}{
  \end{deluxetable*}
}{
  \end{deluxetable}
}
%


\section{Discussion}
\label{sec:discussion}

We present the discovery of \hatcur{},
the first multi-planet system
detected by the HATSouth survey.
The inner planet, \hatcurb{},
is a transiting hot Jupiter
on an eccentric orbit,
completing one revolution every $\approx$ 5 days.
The outer planet, \hatcurc{},
is a cold massive giant planet
on a circular orbit 
with a period of 1422 days.

\subsection{Possible Formation Scenarios of \hatcurbc{} }

The architecture of \hatcurbc{}
poses a challenge for planet formation and migration scenarios. 
Can core accretion explain the 
presence of a hot Jupiter and a massive gas giant 
in the same system?
\cite{Schlaufman:2018} found that planets 
with $M>10 \mjup$ do not preferentially orbit 
metal-rich solar-like stars,
suggesting that these objects most likely 
did not form via core accretion 
but via gravitational instability.
The architecture of \hatcurbc{} hence
suggests that both core accretion 
and gravitational instability
could have occurred in the same system,
which was also previously suggested by 
\cite{Triaud:2017} for WASP-53bc and WASP-81bc.

The current water iceline is around 2.92 au,
suggesting that both
\hatcurb{} and \hatcurc{}
formed beyond the iceline 
and then migrated inwards
to their present locations. 
The presence of \hatcurc{}, 
a massive companion
close to the deuterium burning limit 
\citep{Molliere:2012},
could have scattered \hatcurb{} inwards
resulting in its present eccentric orbit.
Due to its mass,
type-II migration is reduced 
even below the viscous limit for \hatcurc{}
\citep{Baruteau:2014},
resulting in only little inward migration,
potentially explaining its long period.

\subsection{Transit Timing Variations}
\label{sec:ttv}

Variations in the times of transits
can be attributed to the presence of a secondary 
planet in the system 
\citep[e.g.][]{Agol:2005,Mancini:2016,Almenara:2018}.
The maximum transit variation expected 
for the inner planet
is on the order of $10^{-10}$ s,
undetectable with current instruments.
However, this depends on the mutual inclination
between the inner and outer planet.

\subsection{The Inner Transiting Planet \hatcurb{}}

\begin{figure}[]
\plotone{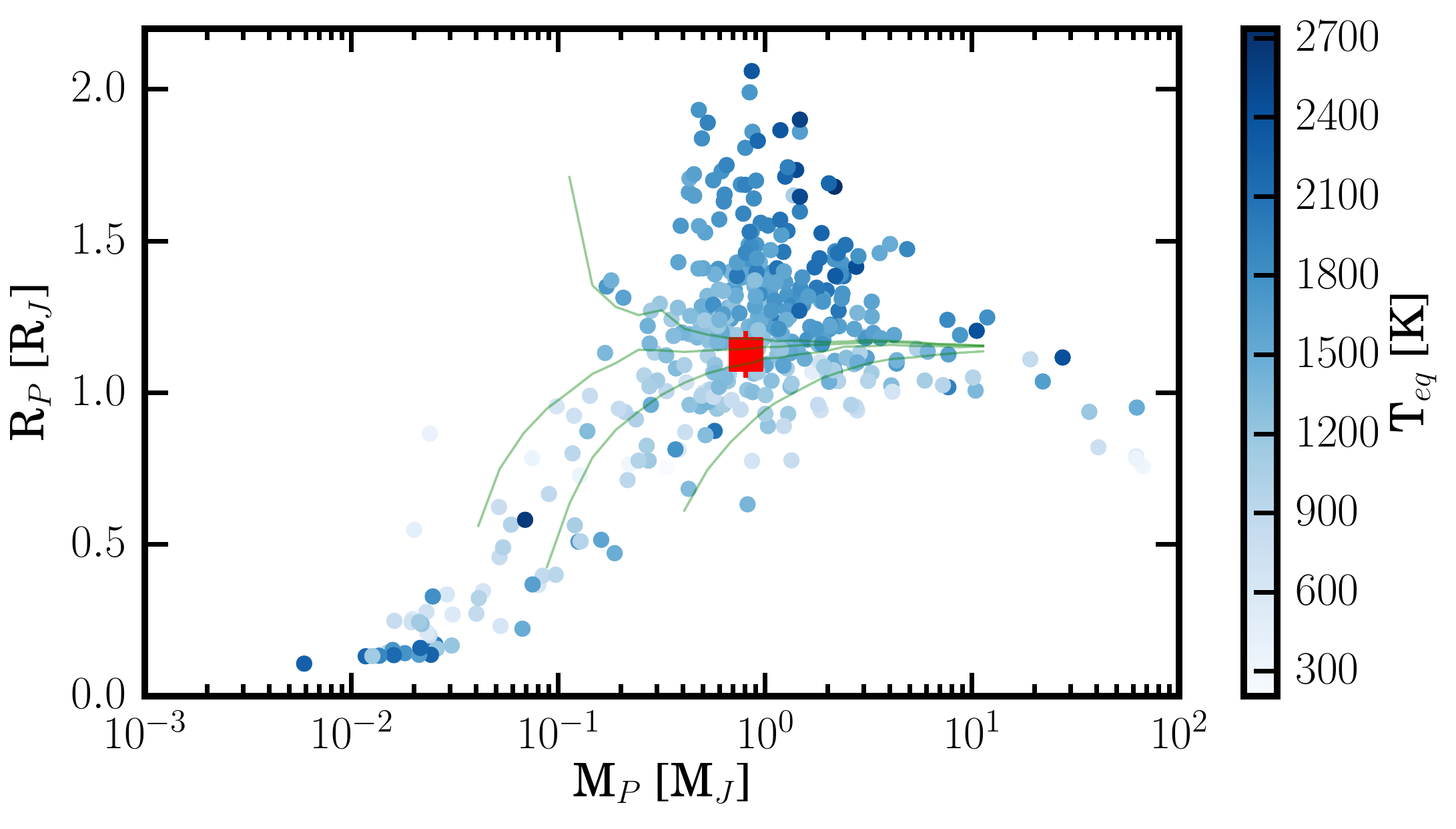}
\caption[]{Mass - Radius diagram for the full population of well characterized transiting planets
colour coded by their equilibrium temperature. \hatcurb{} is shown in red. The
\citet{Fortney:2007} models of planetary structure are also plotted as green lines. The four models
correspond to gaseous planets with $a$=0.045 AU, age=4.3 Gyr, and core masses of 0, 25, 50, and 100 \mearth.
\label{fig:mrdiagram}}
\end{figure}

In Figure \ref{fig:mrdiagram}, we plot the masses and radii of all the transiting exoplanets
having these parameters measured with a precision better than 20\%.
\hatcurb{} lies in a densely populated region of the parameter space, where numerous non
inflated giant planets with similar properties have been detected.
In terms of structure, 
\hatcurb{} is similar to HAT-P-29\,b \citep[$\mpl=0.78 \, \mjup$, $\rpl=1.11 \, \rjup$,
and $P = 5.7$ days;][]{Buchhave:2011}; and K2-115\,b,  \citep[$\mpl=0.84 \, \mjup$,
$\rpl=1.12 \, \rjup$, and $P = 20.3$ days;][]{Shporer:2017}, however with a significantly
shorter period.

We compare the mass and radius of \hatcurb{}  to the theoretical models of
\cite{Fortney:2007}, for a hydrogen-helium dominated planets
with different core masses, at a distance of 0.045 AU, and an age of 4.3 Gyr.
We find that its composition is consistent with a gas-dominated planet 
with a core mass $M_\mathrm{c} < 25 \, \mearth$. However, these
models assume that all the solid material is located inside the core.
According to \citet{th:2016}, \hatcurb{} could have a larger amount
of heavy elements in its interior ($\sim50$ M$_{\oplus}$) if they are 
predominantly mixed in the gaseous envelope.

\begin{figure}[t!]
\plotone{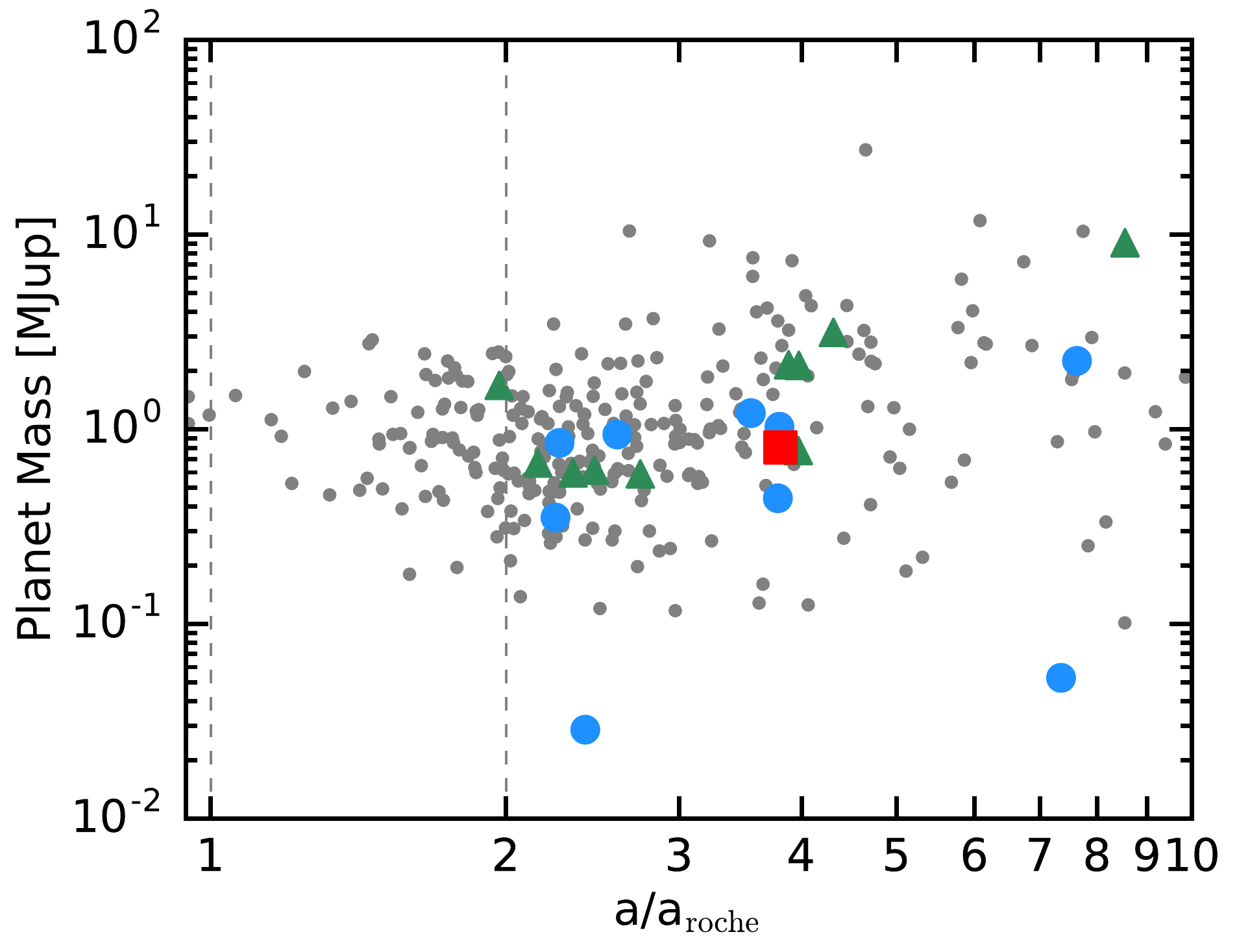}
\caption[]{Planetary mass vs $a/a_\mathrm{roche}$ for single (small gray circles), known multi-planetary systems (blue circles), and systems showing a linear trend (green triangles). \hatcurb{} is shown as a red square. 
Most of the multi-planetary systems 
have $a/a_\mathrm{roche} > 2$, which supports the high eccentricity migration scenario. 
\label{fig:aroche}}
\end{figure}

\begin{figure}[t!]
\plotone{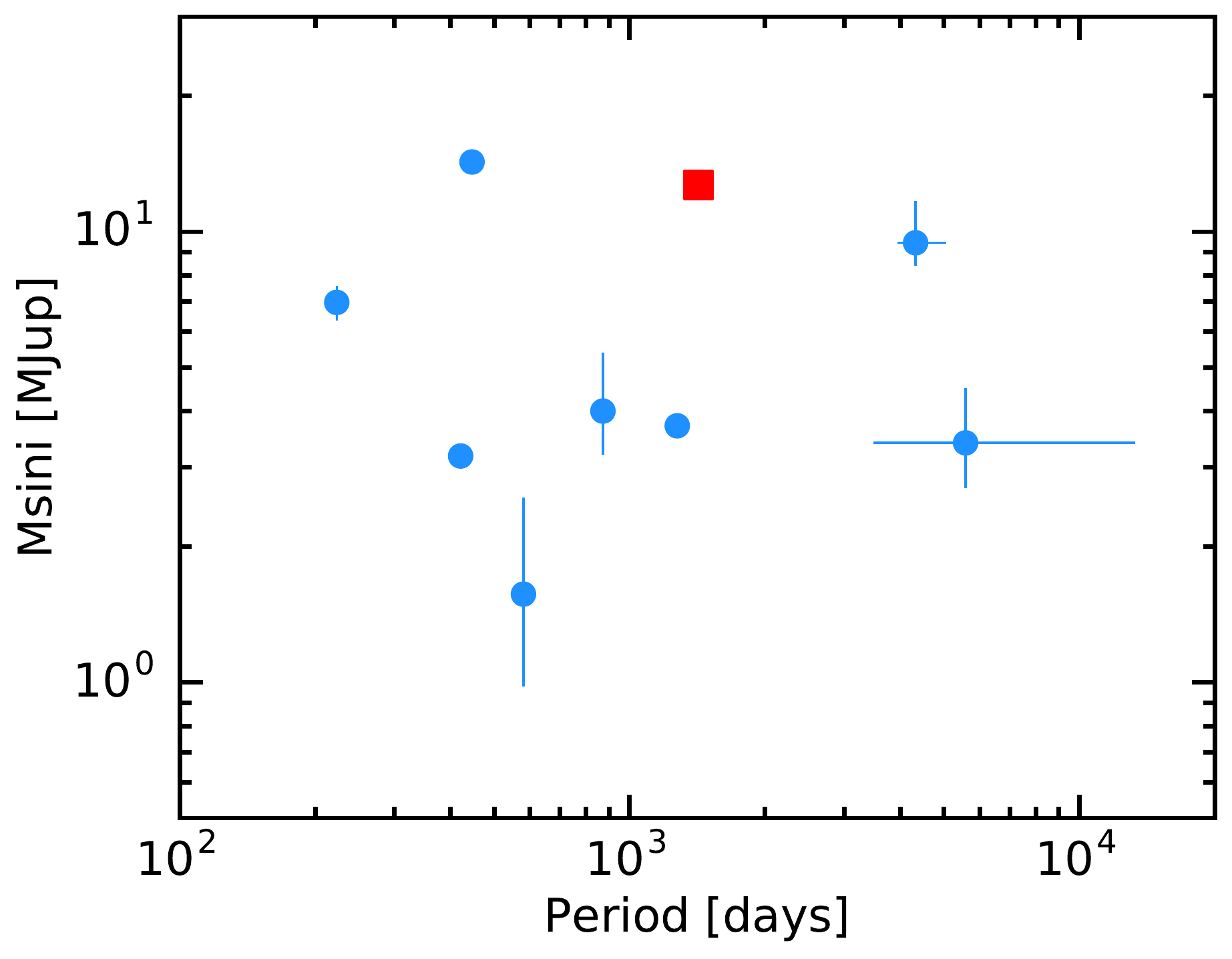}
\caption[]{Msini vs period for the outer companions where the orbit was fully observed (blue circle). 
\hatcurc{} (red square) has the third longest period,
where only 9 companions have been characterised.
\label{fig:companions}}
\end{figure}

\subsubsection{Possible Migration Scenarios of \hatcurb{}}

Hot Jupiters are thought to form beyond the iceline 
and migrate inwards 
via disk or high eccentricity migration,
where the latter requires an outer planetary or stellar companion.
Observations of the projected spin-orbit angle
via the Rossiter-McLaughin (RM) effect 
provides an approach to distinguish between these migration scenarios.
Disk migration predicts circular and aligned orbits,
whereas the high eccentricity migration 
can produce a broad range of obliquities,
depending mostly on the scattering mechanism
and on the effectiveness of tidal interactions
at damping obliquities.

The amplitude of the RM effect scales with $\vsini$, 
the projected rotational velocity of the star.
We predict an RM amplitude of $23-36$ \ms\
for $\vsini = 2.2-3.4$ \kms.
Measuring the RM amplitude for this 
faint star ($V = \hatcurCCtassmv{}$ mag),
is challenging but plausible using HIRES
\citep{vogt:1994,Wang:2018} on the Keck telescope
or with the new high-resolution spectrograph, 
ESPRESSO \citep{Pepe:2014}
at the Very Large Telescope.

Disk migration predicts that planets can migrate up until
they reach the planet-star Roche separation ($a_\mathrm{roche}$),
the critical distance within which a planet 
would start losing mass \citep{Faber:2005}.
On the other hand, high-eccentricity migration predicts 
planets will circularize at a semimajor axis 
greater than $2a_\mathrm{roche}$.
This mechanism would require that hot Jupiters 
are excited to eccentric orbits,
often by being scattered by a distant massive companion,
and survived the tidal dissipation process required to 
circularize their final orbits
\citep{Faber:2005,Ford:2006}.

Many distant planetary companions 
to hot Jupiters have been detected
\citep{Knutson:2014}.
In Figure~\ref{fig:aroche} we
show planetary mass plotted against 
$a/a_\mathrm{roche}$,
where

\begin{equation}
a_\mathrm{roche}  = 2.7 R_p \left( \frac{M_*}{M_p} \right)^{1/3},
\end{equation}

\noindent for all hot Jupiters whose mass and radii are determined
with a precision better than 30\% (small gray circles). 
Blue circles show all the hot Jupiters 
with a fully resolved orbit of the outer planetary companion
and green triangle represent 
the systems whose RVs show a linear trend, 
taken from \citet{Knutson:2014}.
The position of \hatcurb{} is shown with a red square.
All but one multi-planet 
system have $a/a_\mathrm{roche} > 2$, 
HAT-P-7b, 
with a value $a/a_\mathrm{roche}$ 
only slightly lower than 2.
The available data on hot Jupiters with companions 
indicate that high eccentricity migration 
could be the main mechanism for placing 
the gas giant on a close-in orbit 
in these systems.

We compare the parameters of \hatcurc{}
to all the detected planetary companions
whose orbit is fully resolved.
Figure~\ref{fig:companions} shows the position of
\hatcurc{} (red square) on the minimum mass-period diagram
with the other discovered companions (blue circles).
With a period of \hatcurcLCPshort\ days, 
\hatcurc{} has the third longest period,
indicating how few outer companions
to transiting hot Jupiters have been characterised
due to the lack of RV follow-up observations.
All of the companions have minimum masses above 1 \mjup,
which is most likely due to selection effects.

\begin{figure}[t]
\plotone{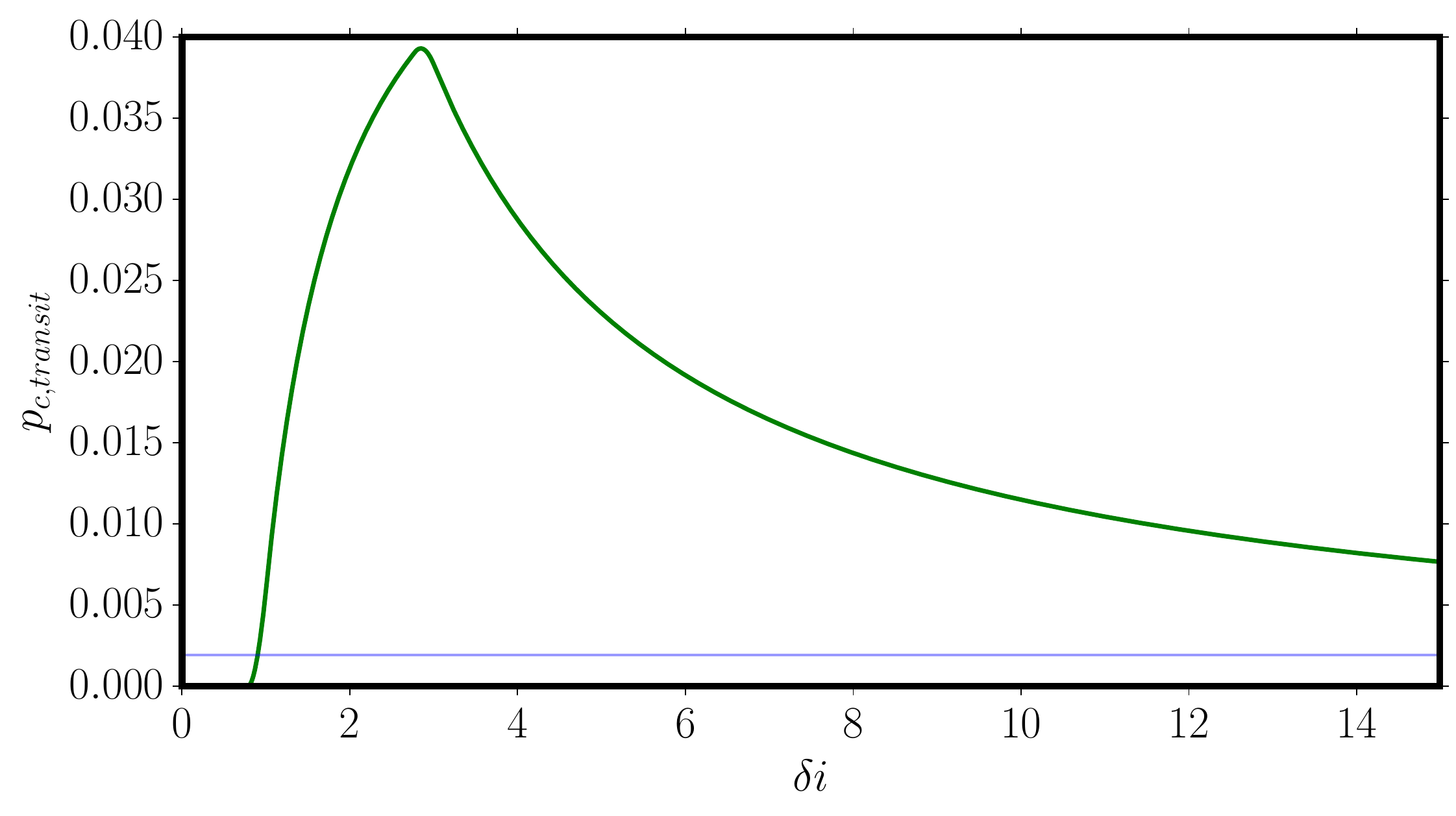}
\caption[]{Transit probability for \hatcurc{} for an aligned configuration with
\hatcurb{} as a function of the maximum separation in inclination between
both planets. The blue line shows the {\em a priori} probability 
for \hatcurc{} to transit.
A maximum probability of $\approx 4$ \% would occur if the 
orbital plane of \hatcurc{}
is inclined around 3 deg with respect to that of \hatcurb{}.
\label{fig:ptrans}}
\end{figure}

\subsection{Possible Transits of \hatcurc{} }

\ifthenelse{\boolean{emulateapj}}{
  \begin{deluxetable}{lc}
}{
  \begin{deluxetable}{lc}
}
\tabletypesize{\scriptsize}
\tablecaption{Future transit windows.
\label{tab:tws}}
\tablehead{
    \multicolumn{1}{c}{Date} &
    \multicolumn{1}{c}{Sun distance} \\
    \multicolumn{1}{c}{(UT)} &
    \multicolumn{1}{c}{(h)}                     
}
\startdata
\noalign{\vskip -3pt}
\\
2021-05-30    & 6.8  \\
2025-04-21    & 9.4   \\
2029-03-13    & 12.2 \\
\enddata

\ifthenelse{\boolean{emulateapj}}{
  \end{deluxetable}
}{
  \end{deluxetable}
}

As was stated in the previous section, knowing the mutual 
inclination between \hatcurb{} and \hatcurc{} can be useful
to further clarify the possible migration path of this system.
The host star is too faint for the GAIA mission to be able to
measure the astrometric signal of \hatcurc{}. 
However, 
the inclination of \hatcurc{} with respect to the plane of the sky 
could be measured if it also transits its star. 
While the {\em a priori} probability
of transit for \hatcurc{} is $\sim0.2\%$, 
if we consider that the two planets are co-planar, 
then the probability of transit
raises by one order of magnitude. 
Figure~\ref{fig:ptrans} shows the
transit probability of \hatcurc\ for different 
assumed maximum mutual inclinations ($\delta_i$) 
between the orbital plane of the planets. 
The probabilities
were computed following the formalism of \citet{Beatty:2010}. 
The maximum probability
(3.8\%) occurs if the mutual inclination 
between the planets is around $3$ $deg$.

The future transit windows for \hatcurc{} are listed in Table \ref{tab:tws}.
In this table we indicate the center of the transit window and the
distance of the target from the Sun at the time of 
putative transit center.
Currently, the
width of the transit window is quite large ($>50$ days) due to the large
uncertainties in the ephemeris. 
Long term RV monitoring of the 
system would be useful to further constrain the width of the transit window.

%


\acknowledgements 
\label{ackno}
\paragraph{Acknowledgements}
Development of the HATSouth
project was funded by NSF MRI grant NSF/AST-0723074, operations have
been supported by NASA grants NNX09AB29G, NNX12AH91H, and NNX17AB61G, and follow-up
observations receive partial support from grant NSF/AST-1108686.
P.S. would like to thank Bertram Bitsch for useful discussions.
A.J.\ acknowledges support from FONDECYT project 1171208, BASAL CATA
PFB-06, and project IC120009 ``Millennium Institute of Astrophysics
(MAS)'' of the Millenium Science Initiative, Chilean Ministry of
Economy. R.B.\ acknowledges support from project
IC120009 ``Millenium Institute of Astrophysics (MAS)'' of the
Millennium Science Initiative, Chilean Ministry of Economy.
L.M. acknowledges support from the Italian Minister of Instruction,
University and Research (MIUR) through FFABR 2017 fund.
J.H.\ acknowledges support from NASA grant NNX14AE87G.
V.S.\ acknowledges support form BASAL CATA PFB-06.  
A.V. is supported by the NSF Graduate Research Fellowship, Grant No. DGE 1144152.
This work has made use of data from the European Space Agency (ESA) mission
{\it Gaia} (\url{https://www.cosmos.esa.int/gaia}), processed by the {\it Gaia}
Data Processing and Analysis Consortium (DPAC,
\url{https://www.cosmos.esa.int/web/gaia/dpac/consortium}). Funding for the DPAC
has been provided by national institutions, in particular the institutions
participating in the {\it Gaia} Multilateral Agreement.
This work is based on observations made with ESO Telescopes at the La
Silla Observatory.
This paper also uses observations obtained with facilities of the Las
Cumbres Observatory Global Telescope.
We acknowledge the use of the AAVSO Photometric All-Sky Survey (APASS),
funded by the Robert Martin Ayers Sciences Fund, and the SIMBAD
database, operated at CDS, Strasbourg, France.
Operations at the MPG~2.2\,m Telescope are jointly performed by the
Max Planck Gesellschaft and the European Southern Observatory.  The
imaging system GROND has been built by the high-energy group of MPE in
collaboration with the LSW Tautenburg and ESO\@.  
We thank the MPG 2.2m telescope support team for their technical 
assistance during observations.

\clearpage
\bibliographystyle{apj}
\bibliography{hatsbib}

\end{document}